\documentclass[twocolumn]{emulateapj}
\usepackage{epstopdf}

\submitted{Version: \today ~~ 
Accepted: February 04, 2013}

\shortauthors{MEN\'{E}NDEZ-DELMESTRE ET AL. 2013}
\shorttitle{OSIRIS VIEW OF SUBMILLIMETER GALAXIES}

\begin{document}

\title{Mapping the Clumpy Structures within Submillimeter Galaxies using Laser-Guide Star Adaptive Optics Spectroscopy} 

\author{\sc Kar\'{i}n Men\'{e}ndez-Delmestre\altaffilmark{1,2,3}, Andrew W. Blain\altaffilmark{2,4}, Mark Swinbank \altaffilmark{5},  Ian Smail\altaffilmark{5}, Rob J. Ivison\altaffilmark{6,7}, Scott C. Chapman\altaffilmark{8}, Thiago S. Gon{\c c}alves\altaffilmark{1}}

\altaffiltext{1}{Observat\'{o}rio do Valongo, Universidade Federal do Rio de Janeiro, Ladeira do Pedro Ant\^{o}nio 43, Rio de Janeiro, RJ 20080-090, Brazil}

\altaffiltext{2}{California Institute of Technology, MC 105-24, Pasadena, CA 91125}

\altaffiltext{3}{NSF Astronomy and Astrophysics Postdoctoral Fellowship; The Observatories of the Carnegie Institution for Science, 813 Santa Barbara St., Pasadena, CA 91101}

\altaffiltext{4}{University of Leicester, University Road, Leicester, LE1 7RH, UK}

\altaffiltext{5}{Institute for Computational Cosmology, Durham University, Durham DH1 3LE, UK}

\altaffiltext{6}{UK Astronomy Technology Centre, Blackford Hill, Edinburgh EH9 3HJ}

\altaffiltext{7}{Institute for Astronomy, Blackford Hill, Edinburgh EH9 3HJ}

\altaffiltext{8}{Institute of Astronomy, Madingley Road, Cambridge, CB3\,0HA, UK}

\email{kmd@astro.ufrj.br}

\begin{abstract}

We present the first integral-field spectroscopic observations of high-redshift submillimeter-selected galaxies (SMGs) using Laser Guide Star Adaptive Optics (LGS-AO). We target H$\alpha$ emission of three SMGs at redshifts $z \sim 1.4-2.4$ with the OH-Suppressing Infrared Imaging Spectrograph (OSIRIS) on Keck.  The spatially-resolved spectroscopy of these galaxies reveals unresolved broad-H$\alpha$ line regions (FWHM$>1000$ km s$^{-1}$) likely associated with an AGN and regions of diffuse star formation traced by narrow-line H$\alpha$ emission (FWHM$\lesssim500$ km s$^{-1}$) dominated by multiple H$\alpha$-bright stellar clumps, each contributing $1-30\%$ of the total clump-integrated H$\alpha$ emission. We find that these SMGs host high star-formation rate surface densities, similar to local extreme sources, such as circumnuclear starbursts and luminous infrared galaxies. However, in contrast to these local environments, SMGs appear to be undergoing such intense activity on significantly larger spatial scales as revealed by extended H$\alpha$ emission over $4-16$\,kpc. H$\alpha$ kinematics show no evidence of ordered global motion as would be found in a disk, but rather large velocity offsets ($\sim~few~\times~100$~km~s$^{-1}$) between the distinct stellar clumps.  Together with the asymmetric distribution of the stellar clumps around the AGN in these objects, it is unlikely that we are unveiling a clumpy disk structure as has been suggested in other high-redshift populations of star-forming galaxies. The SMG clumps in this sample may correspond to remnants of originally independent gas-rich systems that are in the process of merging, hence triggering the ultraluminous SMG phase.

\end{abstract}


%
%

\section{Introduction}

The details of stellar mass assembly are crucial for a full understanding of galaxy evolution.  Most of the present-day stellar mass in massive galaxies was already in place at $z \sim 1$, and the bulk of stellar build-up apparently took place at $1 \lesssim z \lesssim 3$ (e.g., \citealt{dickinson03, perez-gonzalez08}). A population of starbursting galaxies at $z\sim2$ that have been popular candidates to be the progenitors of the most massive galaxies present at $z\sim0$ are the so-called submillimeter galaxies (SMGs; \citealt{blain02}) identified by submm and mm surveys at $\lambda \sim 850 - 125\,\mu$m (\citealt{smail97, hughes98, barger98, barger99, eales99, bertoldi00, cowie02, scott02, borys03, webb03b, coppin05, younger07, weiss09, wardlow11}). With large infrared luminosities (L$_{\rm{8-1000\mu m}} \gtrsim 10^{12}$L$_\odot$) that translate into star-formation rates of SFRs $\gtrsim 100-1000$ M$_\odot$ yr$^{-1}$ \citep{kennicutt98}, SMGs could build the stellar bulk of a massive galaxy in under a few hundred million years \citep{lilly99, genzel03}.

Deep long-slit spectroscopic H$\alpha$ observations of a large sample ($>30$) of SMGs with the Keck instrument NIRSPEC have provided crucial constraints on star formation rates  and dynamical masses of this population (SFR$\sim 1000$ M$_{\odot}$ yr$^{-1}$, M$_{\rm{dyn}}\sim1-2\times~10^{11}$~M$_{\odot}$; \citealt{swinbank04}). These studies reported the presence of broad Balmer line emission (FWHM$_{H\alpha}~\gtrsim~1000$~km~s$^{-1}$) in a large fraction of SMGs ($\gtrsim40\%$), suggesting a direct view to the randomly moving high-velocity gas within the broad-line region of a central active galactic nucleus (AGN). Parallel studies at other wavelengths have revealed a mixture in the astrophysical nature of the underlying power sources in SMGs. Exploiting deep Chandra X-ray observations, \citet{laird10} show that $20-30$\% SMGs host an AGN. However, AGN activity has been shown not to dominate the SMG bolometric luminosities (\citealt{alexander05, alexander08} based on X-ray observations; \citealt{pope08}, \citealt{md09} based on the mid-IR). High-resolution VLBI radio studies of SMGs have also given us insight to the underlying AGN activity in SMGs \citep{biggs10}. 

When both AGN and star formation activity co-exist, long-slit spectroscopic techniques face difficulties in disentangling their independent contributions. Estimates of SFR and dynamical mass based on H$\alpha$ line information in SMGs thus retain the substantial caveat that in the presence of an AGN, the blended nuclear emission may result in the broadening and brightening of the H$\alpha$ emission, potentially leading to AGN-contaminated measurements. Without spatially-resolved information, it is difficult to disentangle the AGN contribution from the nebular emission.

Integral field spectroscopy opens the possibility of investigating the properties of nebular emission at different scales, providing the spatially-resolved information to allow a distinction in spectral properties across a galaxy. Furthermore, considering that dynamical information has proved difficult to extract from long-slit data \citep{swinbank04}, even for the less extreme and likely more ordered cases of optically-selected Lyman-break galaxies (LBGs) at similar redshifts (\citealt{erb03}), an integral spectroscopic insight to the line emission is also the only viable probe of the distribution and dynamics of the gas in the inner galaxy. 

Results from the SPectrometer for Infrared Faint Field Imaging (SPIFFI) on the ESO Very Large Telescope (VLT) first showed how two-dimensional (2D) marginally resolved seeing-limited spectroscopy could already reveal structure in line emission across an SMG at $z\simeq2.5$, SMM\,J14011 +0252  (\citealt{tecza04}). Thereafter, a number of studies using integral field spectrographs at the VLT, Gemini and UKIRT telescopes have explored the spatial distribution and velocity fields of H$\alpha$ emission in a handful of SMGs \citep{swinbank05, swinbank06, nesvadba07, alaghband-zadeh12, harrison12}. Furthermore, high-resolution {\it Hubble Space Telescope} (HST) NICMOS/ACS imaging of 25 SMGs has revealed a mix of morphologies from compact, single-component to clumpy extended structures \citep{swinbank10}, either due to structured dust extinction on a smooth disk distribution or star formation activity concentrated in clumps. Evidence for clumpy star formation is not limited to SMGs but has also been observed in the rest-frame UV/optical continuum of less extreme star-forming galaxies at intermediate and high redshifts (e.g., \citealt{conselice04, elmegreen04a, elmegreen04b, elmegreen05, law07a, overzier10, forster-schreiber11}). Spatially-resolved observations of the ionized gas have provided insight to the underlying kinematics to help discern between a clumpy-disk and a merger scenario (e.g., Keck/OSIRIS: \citealt{stark08}, \citealt{basu-zych09}, \citealt{melbourne09}, \citealt{wright09}, \citealt{law09}, \citealt{goncalves10}, \citealt{jones10}, \citealt{wisnioski12}; VLT/SINFONI: \citealt{forster-schreiber09, genzel11}). 

Recent IFU work by \citet{alaghband-zadeh12} with the GEMINI-NIFS and VLT-SINFONI instruments reveals the predominance of clumpy structure in a sample of 9 SMGs at $2.0<z<2.7$ characterized by multiple galactic-scale subcomponents with an average projected separation of $\sim8$\,kpc and velocity offsets of $\sim200\pm100$ km s$^{-1}$ reminiscent of merging systems. Even higher internal velocities up to $\sim800$ km s$^{-1}$ are identified by \citet{harrison12} in a sample of 8 SMGs at $1.4<z<3.4$, but these are based on [OIII] line emission and attributed to large-scale AGN outflows.

A theoretical framework for the formation of clumpy structures at high redshifts is actively being developed in recent years, with merger (e.g., \citealt{hayward11}) and disk fragmentation simulations successfully reproducing the observational evidence (e.g., \citealt{immeli04, dekel09, bournaud09, bournaud11}). Within the latter scenario, disk fragmentation into giant clumps results from gravitational instabilities triggered by disk-wide turbulence due to rapid gas cooling. The otherwise smooth rotation pattern expected for rotating disks is highly distorted by the velocity offsets of the bright clumps \citep{immeli04}. This ``clumpy stage" is a short-lived phase in the evolutionary scenario of disk galaxies, where disk fragmentation is rapidly followed by clump migration to the center of the galaxy building the central bulge in only a few $\sim10^8$ yr. \citet{dekel09} propose a slightly different scenario where clumpy disks persists for a longer term and are thus more prevalent in the observed population of high-redshift galaxies. Within this scenario cold flows provide a continuous and rapid stream of smooth and clumpy gas along galactic-scale filaments that feed galaxy star formation, keeping the disk gravitationally unstable without destroying it.

In this paper we present the first integral field spectroscopic observations of SMGs aided with Laser Guide Star Adaptive Optics (LGS-AO),  unveiling spatially-resolved details on the clumpy structure in SMGs. We use the OH-Suppressing Infrared Imaging Spectrograph (OSIRIS; \citealt{larkin06}) on the Keck II telescope to study the H$\alpha$ emission in three SMGs in the redshift range $z\sim1.4-2.4$. The superior spatial resolution provided by these observations compared to previous work allows us for the first time to investigate down to kpc-scale detail the internal kinematics and the distribution of spectral properties in these galaxies in the rest-frame optical. We describe our sample selection, observing strategy and the steps comprising the reduction and analysis of the science spectra in \S\ref{obsred}. Our results are presented in \S\ref{results} and discussed in \S\ref{discuss}. We give our conclusions in \S\ref{conclusions}. We assume a $\Lambda$CDM cosmology, with $H_0 = 71$\,km\,s$^{-1}$\,Mpc$^{-1}$, $\Omega_M = 0.27$ and $\Omega_\Lambda = 0.73$.

%
%
\begin{deluxetable*}{llll}
\centering
\tablewidth{0pt}
\tablecolumns{4}
\tabletypesize{\scriptsize}
\tablecaption{Summary of OSIRIS Observations} 
\tablehead{ \colhead{SMM J}  &  \colhead{030227.73} &  \colhead{123549.44} &  \colhead{163650.43}   
\label{Obstab}}
\startdata
\tableline
RA\tablenotemark{a} & 03:02:27.73 & 12:35:49.44 & 16:36:50.43 \\
DEC\tablenotemark{a} & +00:06:53.5 & +62:15:36.8 & +40:57:34.5   \\
$z_{H_\alpha}$\tablenotemark{b} & 1.4076 & 2.2032 & 2.3850  \\
Filter &  $Hn2$ & $Kn2$ & $Kn3$  \\
Wavelength Range\tablenotemark{c} (nm) &  1532.0--1609.5 & 2036.2--2140.7 & 2121.3--2229.4 \\
Plate Scale &  0.1\arcsec & 0.05$\arcsec$ & 0.05$\arcsec$ \\
Exposure  & 12.6 ks & 10.8 ks & 12.6 ks \\
\tableline
\enddata
\tablenotetext{a}{~Radio center from \citet{chapman05}}
\tablenotetext{b}{~From \citet{swinbank04}}
\tablenotetext{c}{~Cut-on and cut-off wavelengths}
\end{deluxetable*}

%
%

\section{Sample Selection, Observations, and Analysis}\label{obsred}

OSIRIS, with spectral resolution $R\simeq 3000-3800$, is a lenslet-based spectrograph that allows for a spatial sampling ranging from $0.02-0.1$\arcsec, depending on the selected lenslet scale. It is designed to be used with the Keck Laser-Guide Star Adaptive Optics (LGS-AO; \citealt{wizinowich06, vandam06}) system. LGS-AO allows for atmospheric distortion corrections to be derived from parallel observing of a spatially coincident laser excited sodium beacon in the upper atmosphere and thus enables close to diffraction-limited resolution in ground-based observations. At the typical redshifts of SMGs, $z \sim 2$ \citep{chapman05}, this corresponds to kpc-scale spatial resolution.

Our targets are from the radio-identified sample of SMGs in \citet{chapman05}.  We selected our science targets carefully to optimize OSIRIS observations, including H$\alpha$ signal-to-noise (S/N), low sky-line contamination and performance of AO-correction. We took advantage of existing near-IR long-slit H$\alpha$ spectroscopy to select SMGs with the brightest H$\alpha$ lines, $S_{H\alpha} \gtrsim 10^{-15}$ erg s$^{-1}$ cm$^{-2}$  \citep{swinbank04, takata06} to maximize detection S/N with OSIRIS. With redshifts in hand we selected SMGs for which the redshifted H$\alpha$ emission does not fall on bright OH sky emission lines, in order to minimize sky contamination. Since the LGS-AO system requires the presence of a relatively-bright ($R \lesssim 18$) star within $1\arcmin$ of the science target in order to derive AO-corrections, we selected SMGs with such nearby $tip$-$tilt$ stars (hereafter, TT-star). The three SMG targets comprising our sample are listed in Table \ref{Obstab}. 

The submm fluxes of the SMGs in our sample ($S_{850\mu m}\sim8$ mJy; \citealt{chapman05}), together with their stellar masses (M$_\star \sim 7\times10^{10}$ M$_\odot$; \citealt{hainline11}) set them well within the range of typical values for the SMG population as a whole. However, we note that by selecting galaxies that are particularly bright in H$\alpha$ compared to the rest of the SMG population, we introduce an important bias favoring SMGs with strong emission lines and likely hosting AGN. Incidentally, all three SMGs in our sample display AGN signatures either in the near-IR, the mid-IR or X-rays. We discuss these individually in Section\,\ref{individual}. 

Our OSIRIS observations include a combination of the $0.05\,\arcsec$ and the $0.1\,\arcsec$ lenslet-scales, which provide the largest fields of view (FOV: $\sim~2.4\arcsec~\times~3.2\arcsec$ and $\sim~4.8\arcsec~\times~6.4\arcsec$, respectively) and also have the advantage of allowing us to maximize the S/N per lenslet, crucial for the study of faint sources.  The science integrations comprise sets of standard ABBA sequences, each comprising four $15-$minute frames taken at two positions separated by $0.5\arcsec$ to allow for on-object dithering.  We followed each of these sets by a dedicated-sky integration frame, offset by $15\arcsec$ from the science target.  In Table \ref{Obstab} we provide details of the observing setup for each of our targets.

To process our data we used the OSIRIS Data Reduction System pipeline\footnote{http://www2.keck.hawaii.edu/inst/osiris/tools/}, which comprises a collection of individual IDL modules, each performing a specific task to ultimately extract each lenslet's spectrum from the {\it raw} 2D data frame and spatially reconstruct the image using a reference map -- a {\it rectification matrix} -- where each pixel within the detector is assigned back to its corresponding lenslet position. The result is a 3D data cube with two spatial axes ($x, y$) and a third axis for wavelength ($\lambda$), calibrated according to the corresponding rectification matrix available for the particular filter and lenslet-scale. 

We improve the sky subtraction by subtracting a supersky 2D image from our science cubes. The supersky image results from median-combining the pipeline-processed sky cubes and collapsing these along the $\lambda$-range around the redshifted H$\alpha$ wavelength ($\Delta\lambda \simeq \lambda_{H\alpha} \pm 600$\,\AA). We also subtract the median flux level within each wavelength channel in the science cubes to eliminate the residual background plateau. Finally, we applied the sky-subtraction routine developed by \citet{davies07}, specifically tailored for near-IR integral-field spectra taking into account the temporal variations in absolute fluxes of OH-lines as well as in their individual fluxes relative to each other. This final step contributed a modest increase in $S/N$ ($\lesssim 1.5$) in our science images.

%
%
\begin{figure*}
\centering
\includegraphics[trim = 0mm 50mm 40mm 50mm, clip, scale=0.63, angle=90]{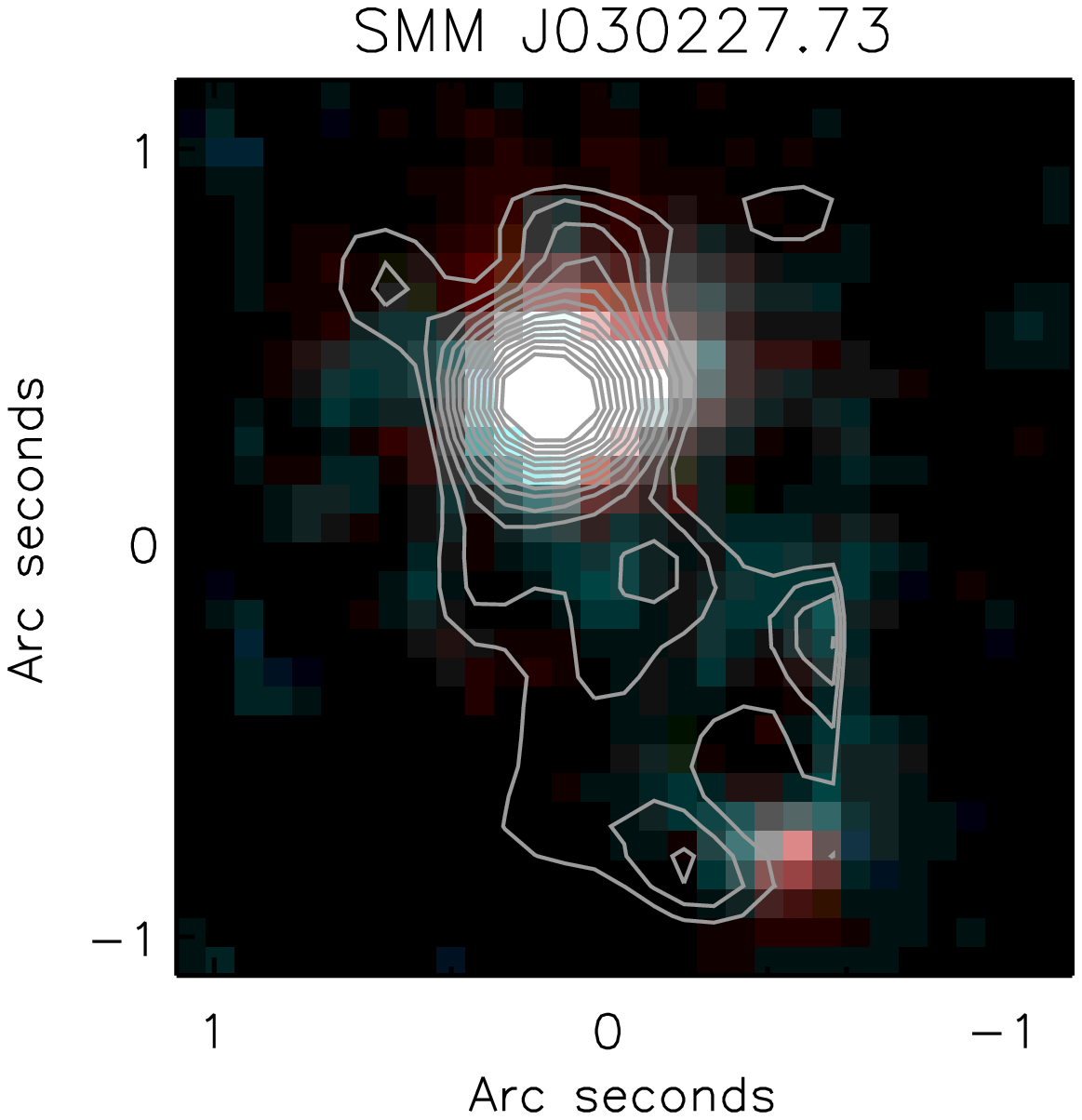}
\includegraphics[trim = 0mm -25mm 0mm 0mm, clip, scale=0.3, angle=0]{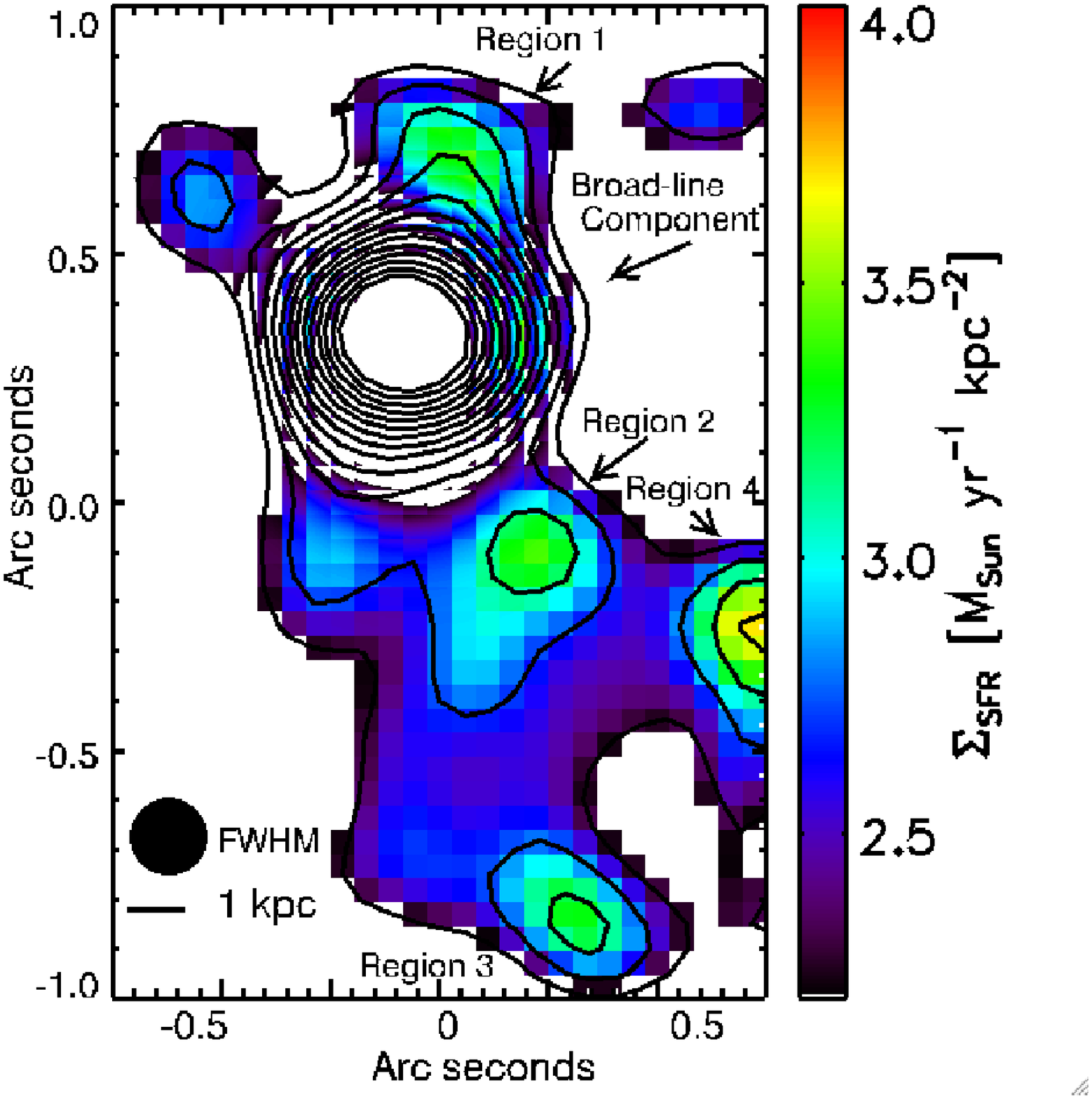}
\caption{\small Distribution of H$\alpha$ emission in SMM\,J030227.73, where North direction is up and East is left. ({\bf Left:})~OSIRIS H$\alpha$ intensity contour map (inner/outer surface brightness contours corresponding to $\sim 50/ 10\sigma$) overlaid on a multicolor HST-WFPC2/NICMOS (B-/H-band) image. H$\alpha$ emission is dominated by a compact knot, with diffuse  line emission stretching towards the red continuum-emitting knot in the south-western direction. We note that the OSIRIS field of view does not cover the full extent of the background image, but stops short of proper coverage of the red knot. ({\bf Right:}) Flux-calibrated map of (projected) SFR surface density prior to any extinction correction, where the broad-line H$\alpha$ emission has been subtracted (see \S\ref{maps} for details). Note that the H$\alpha$ emission is concentrated in multiple clumps. Labels indicate distinct H$\alpha$-bright regions that dominate the diffuse H$\alpha$ emission. We study these regions individually. The solid circle on the bottom left corner is the FWHM of the TT star taken just prior to the science exposures to represent the PSF of the observations; the horizontal bar represents  1\,kpc at the redshift of the target. 
\label{c3_15maps1}}
\end{figure*}

%
%

\subsection{H$\alpha$ Maps}\label{maps}

\begin{figure*}
\centering
\includegraphics[trim = 0mm 50mm 0mm 5mm, clip, scale=0.5, angle=90]{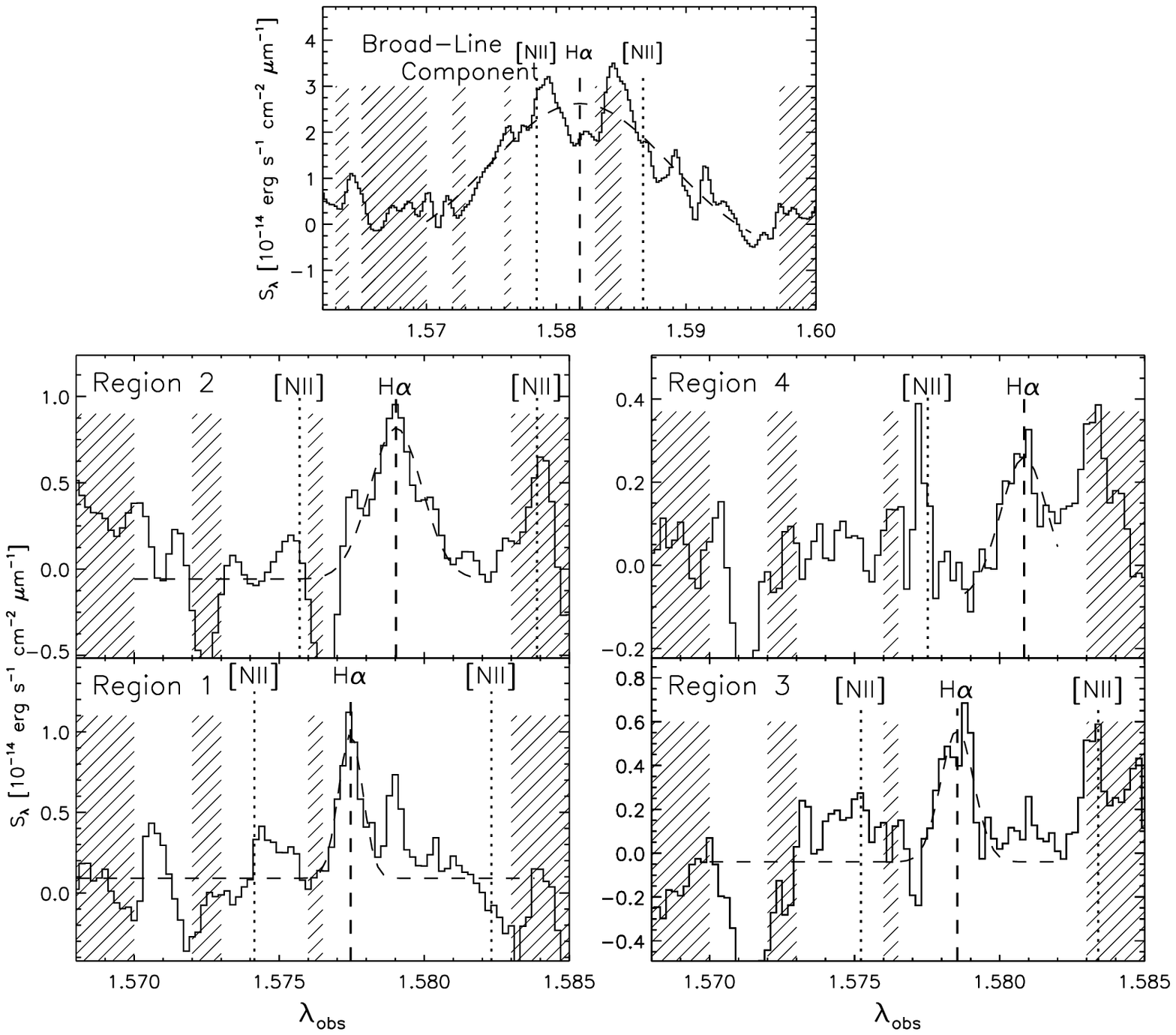}
\includegraphics[trim = -50mm -25mm 0mm 0mm, clip, scale=0.3, angle=0]{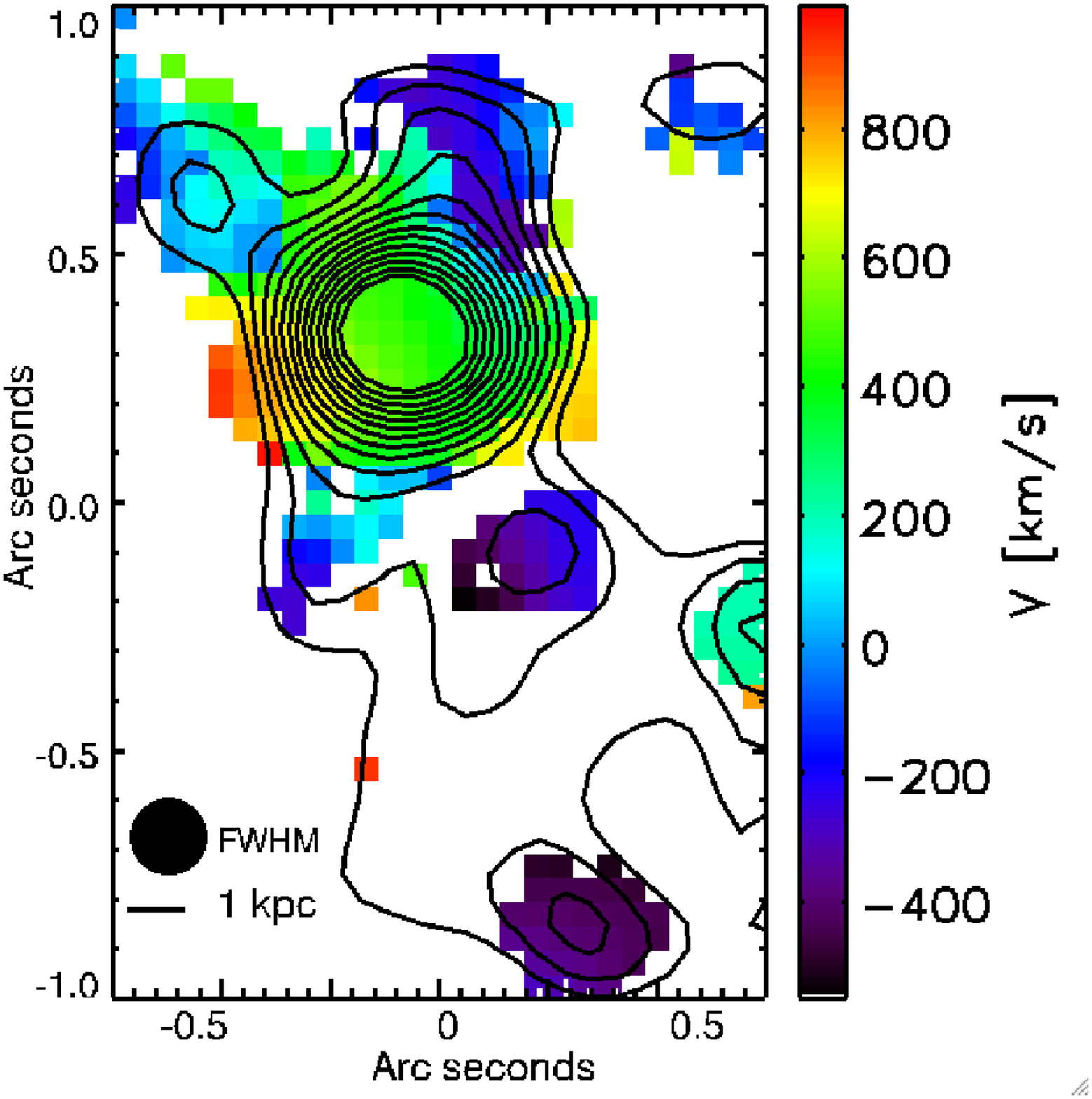}
\includegraphics[trim = 0mm -25mm 0mm 0mm, clip, scale=0.3, angle=0]{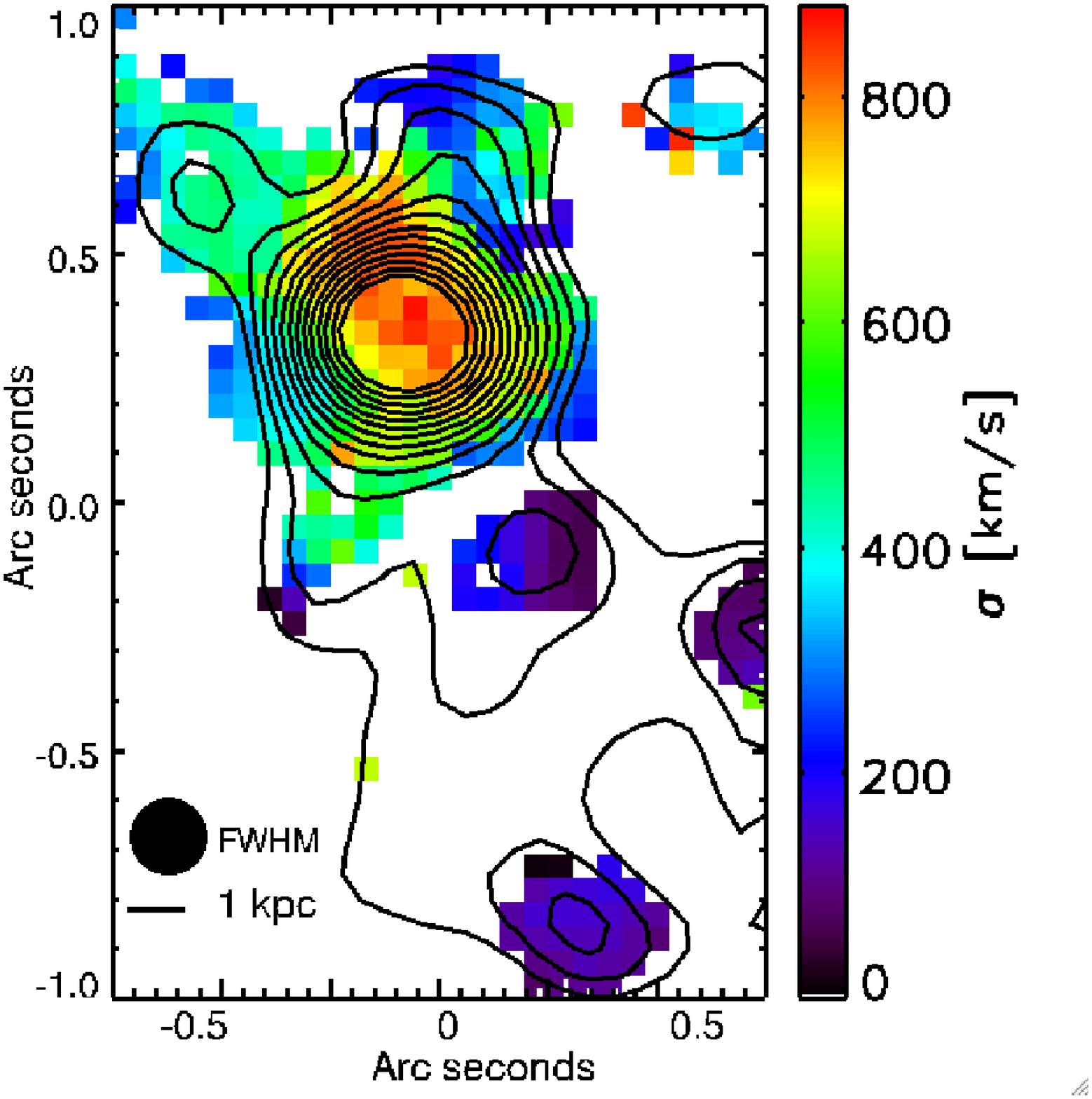}
\caption{\small Spectral properties and kinematic maps for SMM\,J030227.73. ({\bf Top:}) H$\alpha$ 1D spectra extracted from individual H$\alpha$-bright regions (see Fig.\,\ref{c3_15maps1}): the broad-H$\alpha$ component corresponds to the dominating point source and the 4 narrow H$\alpha$ components are associated to distinct clumps, one adjacent to the broad line region and 3 within the diffuse region stretching in the South-West direction. The location of bright OH lines are shown by  diagonally-hashed vertical columns. ({\bf Bottom Left:})  H$\alpha$ velocity map revealing no global velocity gradient across the galaxy, but velocity offsets ($\sim 200-600$ km s$^{-1}$) between the bright point-source component and the other H$\alpha$-bright clumps. ({\bf Bottom Right:}) H$\alpha$ line-width ($\sigma_{H\alpha}$) map showing the point-source component with broad-H$\alpha$ emission relative to that of the remaining clumps. We attribute the broad H$\alpha$ emission to AGN activity, while the emission within the remaining clumps and diffuse region is more likely associated with star-formation. 
\label{c3_15maps2}}
\end{figure*}

We construct H$\alpha$ intensity, velocity ($v_{\rm{H\alpha}}$) and line-width ($\sigma_{\rm{H\alpha}}$) maps for each target to help us understand the distribution of H$\alpha$ properties across the SMGs in our sample (see Figs.\,\ref{c3_15maps1}-\ref{16t3maps2}). Following a similar approach to \citet{law09} and \citet{goncalves10}, we convolve the individual image frames at each wavelength channel with a 2D gaussian kernel (FWHM\,$=1.5$ pixels) in an effort to improve the S/N  while remaining below the spatial resolution limit of our observations as determined by our PSF (FWHM$\sim0.10-0.25$\arcsec), measured from the TT-star peak up imaging. For SMM J123549.44 and SMM J163650.43 we also smooth the one-dimensional (1D) spectra at each spatial pixel with a 2\AA- and 5\AA-wide gaussian kernel, respectively, to better distinguish the line emission from the underlying noise. 

Imposing a set of line-search thresholds for H$\alpha$ line-width ($\lesssim 2500$ km s$^{-1}$), S/N with respect to an off-source region ($\gtrsim 6$) and velocity offset relative to the long-slit H$\alpha$ redshift ($\lesssim 1000$ km s$^{-1}$) for our targets from \citet{swinbank04}, we fit single Gaussian profiles to fit the H$\alpha$ line emission along the dispersion direction at each spatial pixel. We use the collection of these Gaussian fits, in particular the central wavelength and the line width (corrected for instrumental width) at each spatial pixel, to construct an H$\alpha$ velocity map, reflecting the relative spectral position of the line centroids, and an H$\alpha$ line-width map characterizing the velocity dispersions of the H$\alpha$ emission. Note that only the Gaussian fits complying with our imposed thresholds are used for the velocity and line-width maps, while other ones are disregarded (i.e., empty pixel in the resulting maps). The selection of these S/N, velocity and line-width thresholds were not established a priori, but tuned carefully to each object, in order to eliminate spurious elements from the resulting maps. We produce H$\alpha$ intensity maps by collapsing the final science data cube along the dispersion direction, stacking the science data cube's flux over a range $\Delta v \simeq 1000-2000$ km s$^{-1}$ centered at the redshifted H$\alpha$ line. In Figs.\,\ref{c3_15maps2}, \ref{12t1maps2}, \ref{16t3maps2} we show H$\alpha$ velocity and line-width maps with overlaid H$\alpha$ intensity contours for SMM J030227.73, SMM J123549.44, and SMM J163650.43, respectively. We oversample these maps by a factor of 2 to improve their visual appearance. We also show the 1D spectra integrated over several kpc-sized H$\alpha$-bright regions, which are labeled in the maps shown in Figs.\,\ref{c3_15maps1}, \ref{12t1maps1}, \ref{16t3maps1}.

We flux calibrate the H$\alpha$ line emission in our OSIRIS observations using as a reference the H$\alpha$ line fluxes from long-slit spectroscopy by \citet{swinbank04}, where slit loss correction has been performed based on $K$-band images.  We set the galaxy-integrated long-slit H$\alpha$ fluxes equal to the integrated OSIRIS flux resulting from collapse of the science cube along the dispersion direction ($\Delta v \simeq1000-2000$ km s$^{-1}$) centered at the redshifted H$\alpha$ line. By adopting this approach, we ensure that we are not artificially boosting the H$\alpha$ emission within the brightest clump regions.  In principle, peak-up imaging of the TT-star for each science target is also a viable reference to perform absolute flux calibrations. However, we note that when using the AO technique, a large flux uncertainty arises from the imperfect channeling of light into the AO-corrected PSF. Following detailed modeling of this issue by  \citet{law07b}, the total flux uncertainty for the reference TT-star flux corresponds to {\it at least} $\sim30$\%. Furthermore, because flux calibration requires also taking into account the spatial offset between the science target and the TT-star to consider the {\it true} PSF at the science target location, the uncertainties in absolute flux calibration based on TT-star observations increase significantly.

In order to disentangle the diffuse H$\alpha$ component ($\sim0.5-2\arcsec$) from the compact one (FWHM $\sim0.2-0.4\arcsec$), we rely on a target-specific PSF determined from peak-up imaging of the TT-star to subtract the compact region of high-S/N line emission from each map. This is particularly important considering the potential spread of emission beyond the AO-corrected PSF that may escape the immediate regions of these high-S/N regions of emission. In this way we avoid potential contamination and ensure the best separation of the diffuse and compact H$\alpha$ emission components. The PSF-subtracted images that we use to investigate diffuse emission are shown on Figs.\,\ref{c3_15maps1}, \ref{12t1maps1}, and \ref{16t3maps1} for SMM J030227.73, SMM J123549.44, and SMM J163650.43, respectively.

%
%

%
%

\section{Results}\label{results}

\begin{figure*}
\centering
\includegraphics[trim = 0mm 50mm 40mm 50mm, clip, scale=0.5, angle=90]{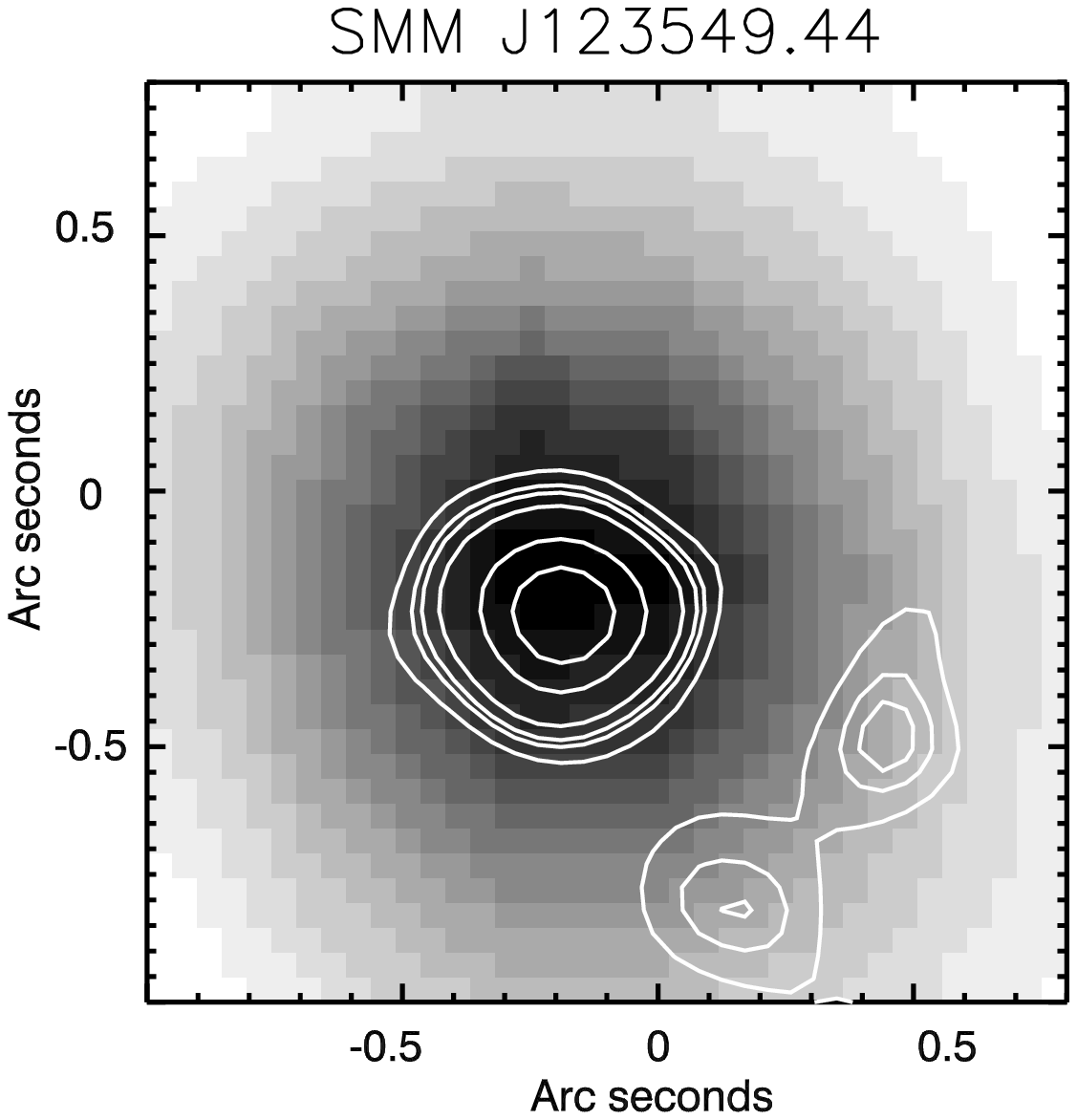}
\includegraphics[trim = 0mm -20mm -20mm 0mm, clip, scale=0.27, angle=0]{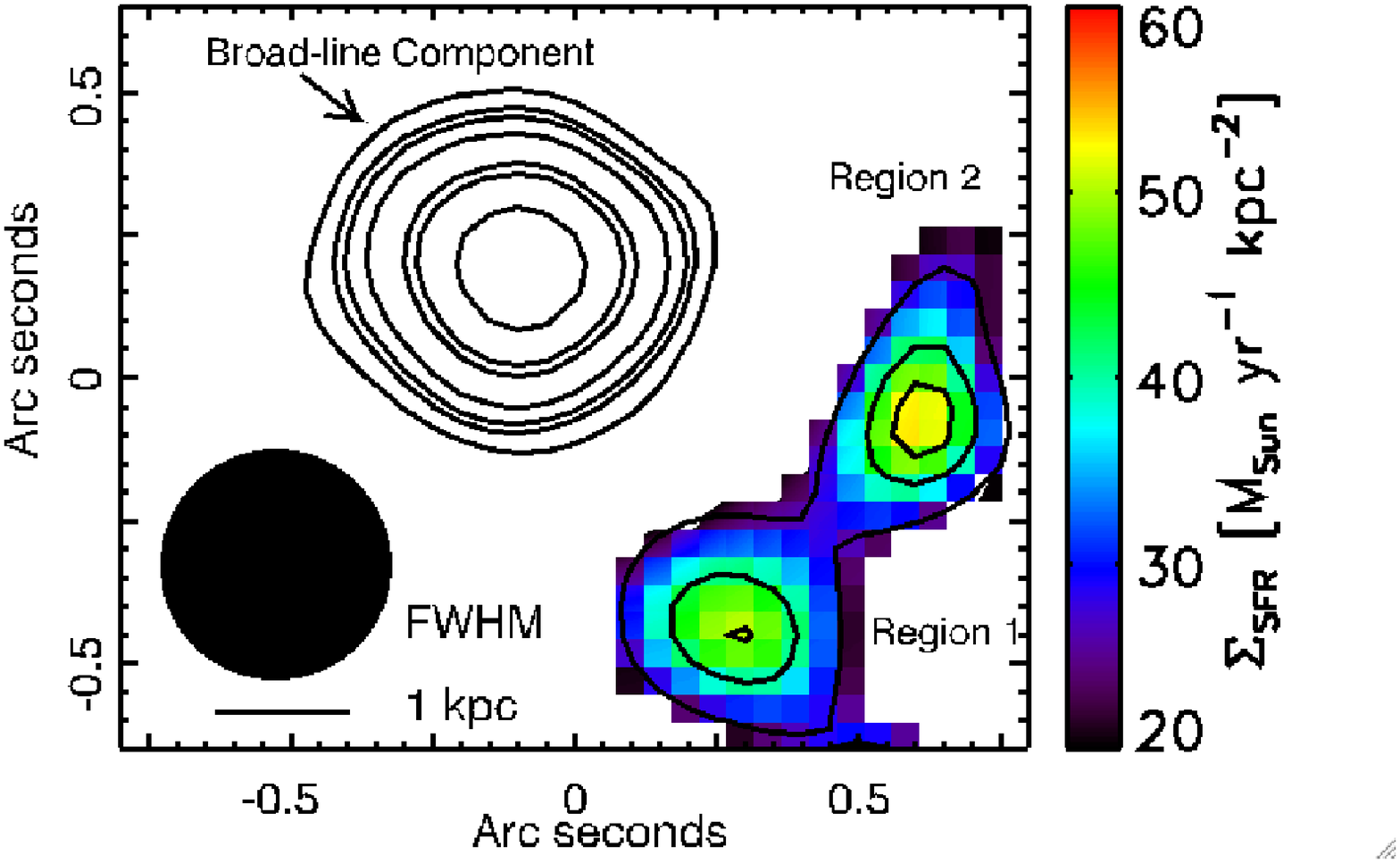}
\caption{Distribution of H$\alpha$ emission in SMM\,J123549.44, following the same format as Fig.\,\ref{c3_15maps1}. The H$\alpha$ OSIRIS contours (inner/outer surface brightness contours corresponding to $\sim 40/ 10\sigma$) are overlaid on Subaru $i$-band imaging from the Hawaii Hubble Deep Field North. The H$\alpha$ emission is dominated by a compact source, with two fainter H$\alpha$ clumps roughly $\sim0.4\arcsec$ the South-West.
\label{12t1maps1}}
\end{figure*}

The OSIRIS H$\alpha$ intensity maps are shown in Figs.\,\ref{c3_15maps1}, \ref{12t1maps1}, \ref{16t3maps1} as contours overlaid on near-IR images tracing the rest-frame optical continuum of these galaxies. The H$\alpha$ contours allow us to map the overall spatial distribution of the line emission and to distinguish between unresolved kpc-sized compact regions of high S/N emission and a more diffuse component, where the former are likely associated with nuclear activity, either an AGN or a compact starburst, and the latter corresponds to star formation. Note the prevalence of clumpy structure in the sources on 1-2 kpc scales. The line-width and velocity maps for each of our targets in Figures \ref{c3_15maps2}, \ref{12t1maps2} and \ref{16t3maps2} allow us to probe for velocity offsets between distinct components in each system and for the presence of AGNs, as revealed by large line-widths (FWHM $ > 1000$ km s$^{-1}$). From these OSIRIS observations we derive clump-integrated velocity dispersions ($\sigma_{\rm{H\alpha}}$), star-formation rates (SFRs) and star-formation rate surface densities ($\Sigma_{\rm{SFR}}$). We list these in Table\,\ref{ResultsTab}.

\bigskip
\bigskip
%
%

\subsection{Notes on individual SMGs}\label{individual}

%
%

%
%

\subsubsection{SMM\,J030227.73}\label{results_c3_15}

SMM\,J030227.73 (aka CUDSS\,3.15) was first detected and identified by \citet{webb03a}. Optical and near-IR spectra by \citet{chapman05} and \citet{swinbank04} place this galaxy at $z=1.408$, with a high [NII]/H$\alpha$ ratio ([NII]/H$\alpha~=~1.38 \pm~0.07$; \citealt{swinbank04}) suggesting the presence of AGN activity. \citet{harrison12} also identify spatially-unresolved ($\le3.6$ kpc), broad emission (FWHM$=900\pm300$ km $s^{-1}$) in galaxy-integrated [OIII] emission for this object. Furthermore, $Spitzer$ mid-IR spectroscopic follow-up of this object shows a substantial excess in mid-IR continuum, suggesting a significant AGN contribution (\citealt{md09}; see also \citealt{hainline09}). The near-IR continuum map of SMM\,J030227.73 in Fig.\,\ref{c3_15maps1} shows that continuum emission is dominated by a bright, compact source, with a much fainter secondary component $\sim1.3\arcsec$ ($\sim11$ kpc) to the South-West. With a high [NII]/H$\alpha$ ratio, this secondary component has been tentatively identified with an AGN by \citet{swinbank06}. A faint H$\alpha$-bright {\it bridge} was also detected between the two point-source components in the non-AO H$\alpha$ observations of \citet{swinbank06}. 

The OSIRIS H$\alpha$ contours in Fig.\,\ref{c3_15maps1} show a compact peak of H$\alpha$ emission (FWHM$\simeq 0.36\arcsec$, corresponding to $\sim3$ kpc) that coincides with the continuum-bright compact region. Although the South-West secondary continuum-detected compact source lies just outside our field of view, OSIRIS observations allow us to spatially resolve the bridge region between the two continuum-detected compact components. As shown in Fig.\,\ref{c3_15maps1} H$\alpha$ contours highlight the bridge region of diffuse H$\alpha$ emission extending out to $\gtrsim 1\arcsec$ (8.5 kpc) to the South-West of the central bright source. In Fig.\,\ref{c3_15maps2}, we show the H$\alpha$ line-width ($\sigma_{H\alpha}$) and velocity maps for SMM\,J030227.73 with overlaid H$\alpha$ flux contours. We also show the 1D H$\alpha$ spectra extracted from distinct spatial regions:  the bright point source and 4 distinct clumps within the diffuse H$\alpha$ emission, 3 of which lie in the H$\alpha$-bright bridge region extending to the South-West.  The bright, compact knot from the central source stands out at a $S/N \sim 40$ with broad H$\alpha$ emission ($\sigma_{rest} \gtrsim1000$ km s$^{-1}$; see 1D spectrum in Fig.\,\ref{c3_15maps2}), while to the South-West direction fainter and diffuse narrow H$\alpha$ emission appears at a lower, but still significant $S/N \sim 10-20$ ($\sigma_{rest} \lesssim 400$ km s$^{-1}$). The spatial distribution of H$\alpha$ line widths conforms with previous observations by \citet{swinbank06} where the bright central source was identified as an AGN. However, since the spatial coverage of our observations does not properly extend out to the South-Western knot shown in Fig.\,\ref{c3_15maps1} (the shown contours are at the very edge of the OSIRIS FOV), we cannot confirm the possibility of two AGNs within this system, as suggested by \citet{swinbank06}. 

The velocity map in Fig.\,\ref{c3_15maps2} suggests a kinematically-rich structure, where the H$\alpha$ clumps to the South-West are offset from the compact component by $\sim 200-600$ km s$^{-1}$. These velocity offsets are distinct from the velocity found by \citet{swinbank06} for the secondary compact clump that they associate with a secondary AGN in the system ($v \sim 90 \pm 20$ km s$^{-1}$). This suggests that the extended H$\alpha$ emission is not directly associated with the secondary continuum source to the South-West of the bright point source. 

The H$\alpha$ arising from the OSIRIS-detected bright point source accounts for $\sim90$\% of the clump-integrated H$\alpha$ observed in the diffuse region. We note that this contribution from the AGN to the H$\alpha$ emission is an upper limit because: (1) we clearly do not cover the full extent of the SMG, (2) are not sensitive to fainter emission and (3) have no constraints on the potential spatially-inhomogeneous dust extinction that has been shown to be significant in SMGs \citep{swinbank10}.

%
%
\begin{figure*}
\centering
\includegraphics[trim = 0mm 0mm 40mm 0mm, clip,scale=0.5, angle=90]{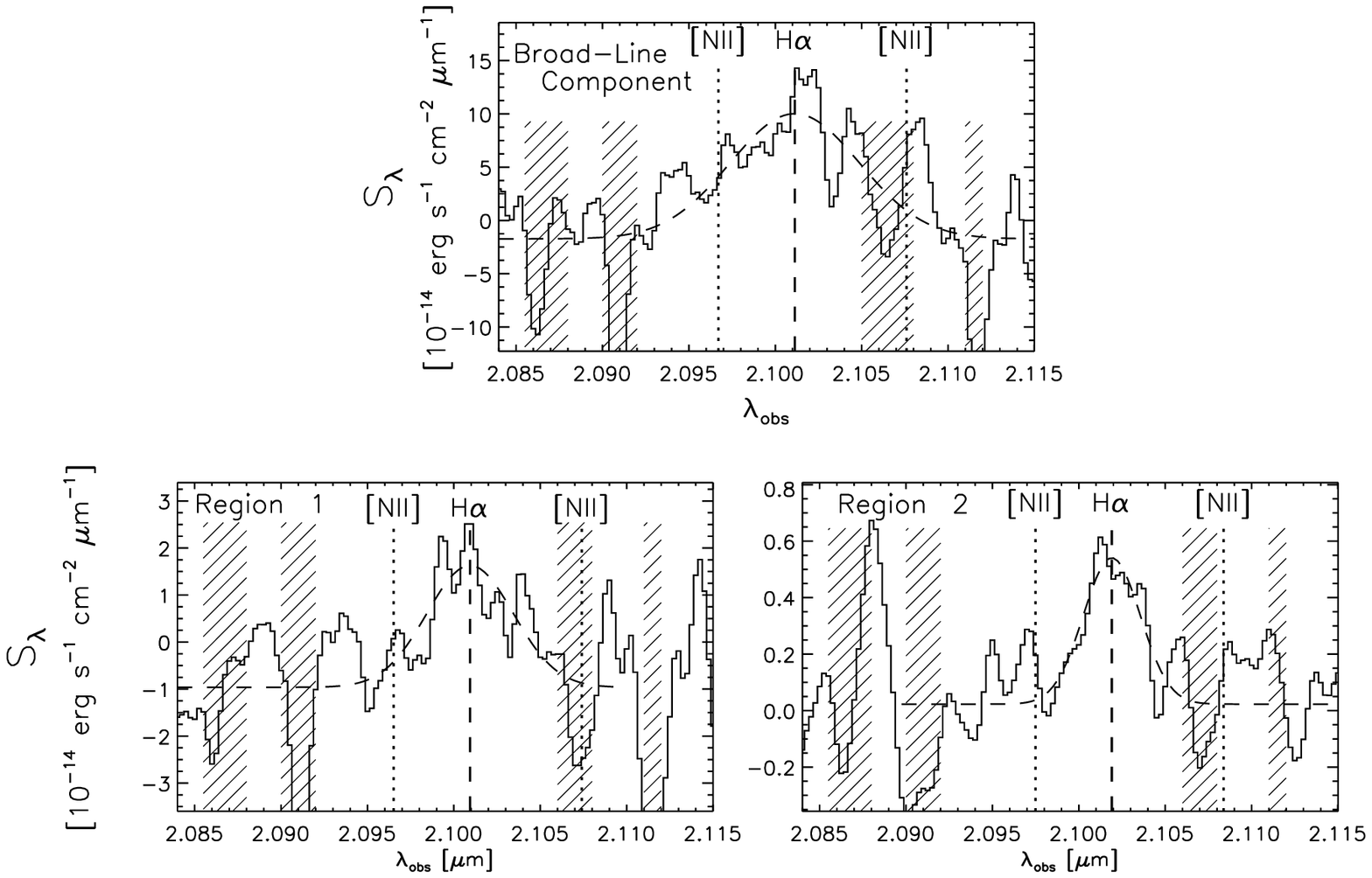}
\includegraphics[scale=0.27, angle=0]{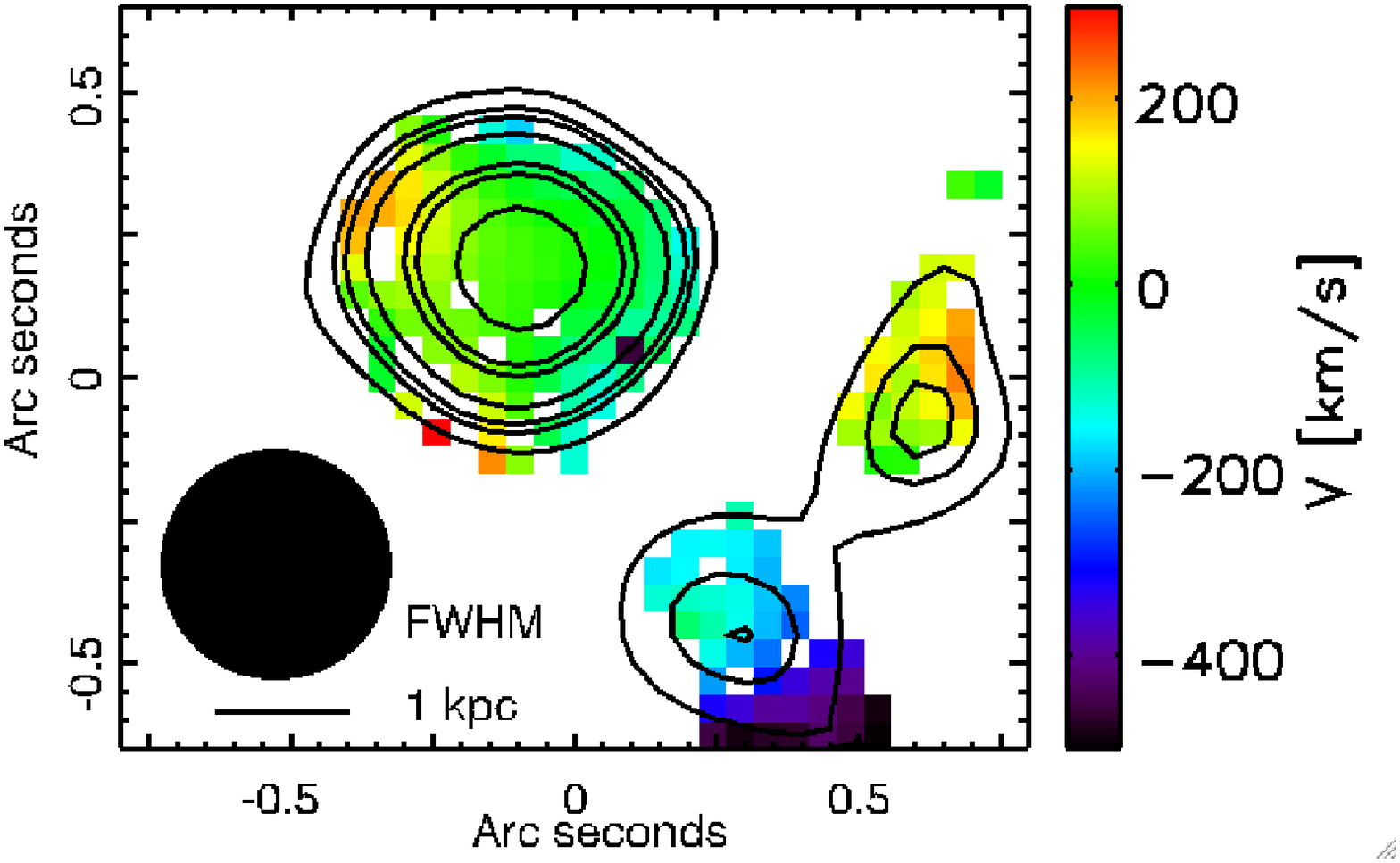}
\includegraphics[scale=0.27, angle=0]{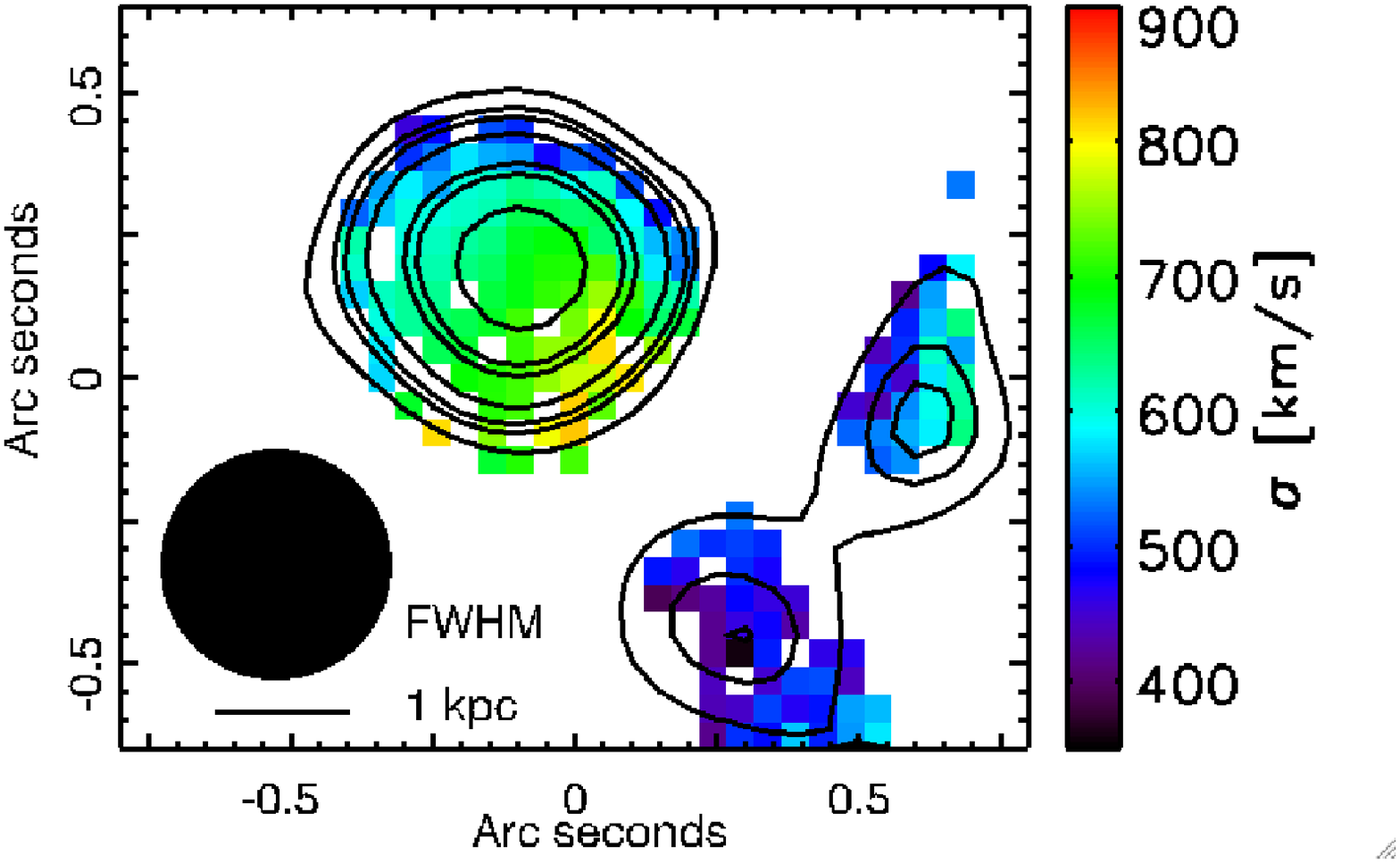}
\caption{H$\alpha$ kinematics in SMM\,J123549.44, following the same format as Fig.\,\ref{c3_15maps2}. The bright point-source emission that dominates the H$\alpha$ intensity map displays broad H$\alpha$ emission at a redshift $z=2.201$ and is likely associated with an AGN. At a fainter S/N, the two distinct components to the South-West (see Fig.\,\ref{12t1maps1}) show narrower H$\alpha$ line emission and are likely dominated by star formation activity (see Table\,\ref{ResultsTab}).
\label{12t1maps2}}
\end{figure*} 

%
%

\subsubsection{SMM\,J123549.44}

Ultra-deep Chandra observations of this source have revealed clear AGN signatures (\citealt{alexander05}). Furthermore, recent mid-IR spectroscopic \citep{md09} and photometric observations \citep{hainline11} have revealed an excess of hot dust continuum emission, in agreement with the hypothesis of a significant contribution from AGN activity to the bolometric luminosity. \citet{tacconi06, tacconi08} have also undertaken high-resolution CO observations of this galaxy and find CO emission dominated by a compact source ($\lesssim 0.5\arcsec$) with a prominent double-peaked CO profile, which they associate with the orbital motions of gas within a disk close to an AGN.

Optical and near-IR (long-slit) spectroscopy set this object at $z=2.203$ \citep{chapman05, swinbank04}. Fig.\,\ref{12t1maps1} shows that the H$\alpha$ emission for SMM\,J123549.44 is dominated by a compact source with FWHM $\simeq0.24\arcsec$ (3 kpc), though faint traces of diffuse emission extend out to $0.5\arcsec$ ($\sim4$ kpc) to the South-West of the central region.  From long-slit observations, \citet{swinbank04} report a relatively narrow H$\alpha$ emission of FWHM$_{rest}=540\pm30$~km~s$^{-1}$. In contrast, \citet{harrison12} identify in the galaxy-wide [OIII] spectrum an extremely broad-line emission (FWHM$=1500$ km s$^{-1}$) attributed to a possible AGN outflow, while a hint of extended [OIII] emission ($\lesssim0.8\arcsec$) is reported with S/N insufficient for proper constraints on its spectral properties.

Our observations unveil the H$\alpha$ line-width distribution shown in Fig.\,\ref{12t1maps2}, allowing us to clearly distinguish broad H$\alpha$ line emission arising from a bright compact region, with FWHM $=1260\pm210$ km s$^{-1}$ and a velocity offset of $\sim-150\pm50$ km s$^{-1}$ from the redshift reported by \citet{swinbank04}. Narrower H$\alpha$ emission (FWHM$\sim 500-800$ km s$^{-1}$) characterizes the two fainter H$\alpha$ clumps to the South-West (see Fig.\,\ref{12t1maps2}, Table\,\ref{ResultsTab}).   The contribution from the broad H$\alpha$ region to the total emission within the H$\alpha$-bright clumps amounts to $\sim90$\%, keeping in mind similar caveats stated for SMM\,J030227.73. Adopting the broad-line redshift as that of the system, we report a redshift of $z=2.2012$. 

The velocity map shown in Fig.\,\ref{12t1maps2} suggests internal velocity offsets of $\sim100-200$~km~s$^{-1}$, consistent with the velocities reported by \citet{tacconi08}. Although these velocities could reflect the ordered motions of a gaseous disk, resulting in a total dynamical mass M$_{\rm{dyn}}=\sigma^2R/G\sim4\times10^{10}$M$_\odot$, the asymmetrical disposition of the narrow-line H$\alpha$-bright clumps around the broad-line region suggests that this is not a well-behaved disk structure with an AGN at its center. Hence, the velocity offsets could also be attributed to the relative motion of multiple galactic-scale companions, potentially undergoing a merger. However, the low $S/N$ of the southern and western components and the potential for an underlying lower surface-brightness plane of emission below the instrument's sensitivities remain an important caveat.

%
%
\begin{figure*}
\centering
\includegraphics[trim = 0mm 75mm 0mm 53mm, clip, scale=0.65, angle=90]{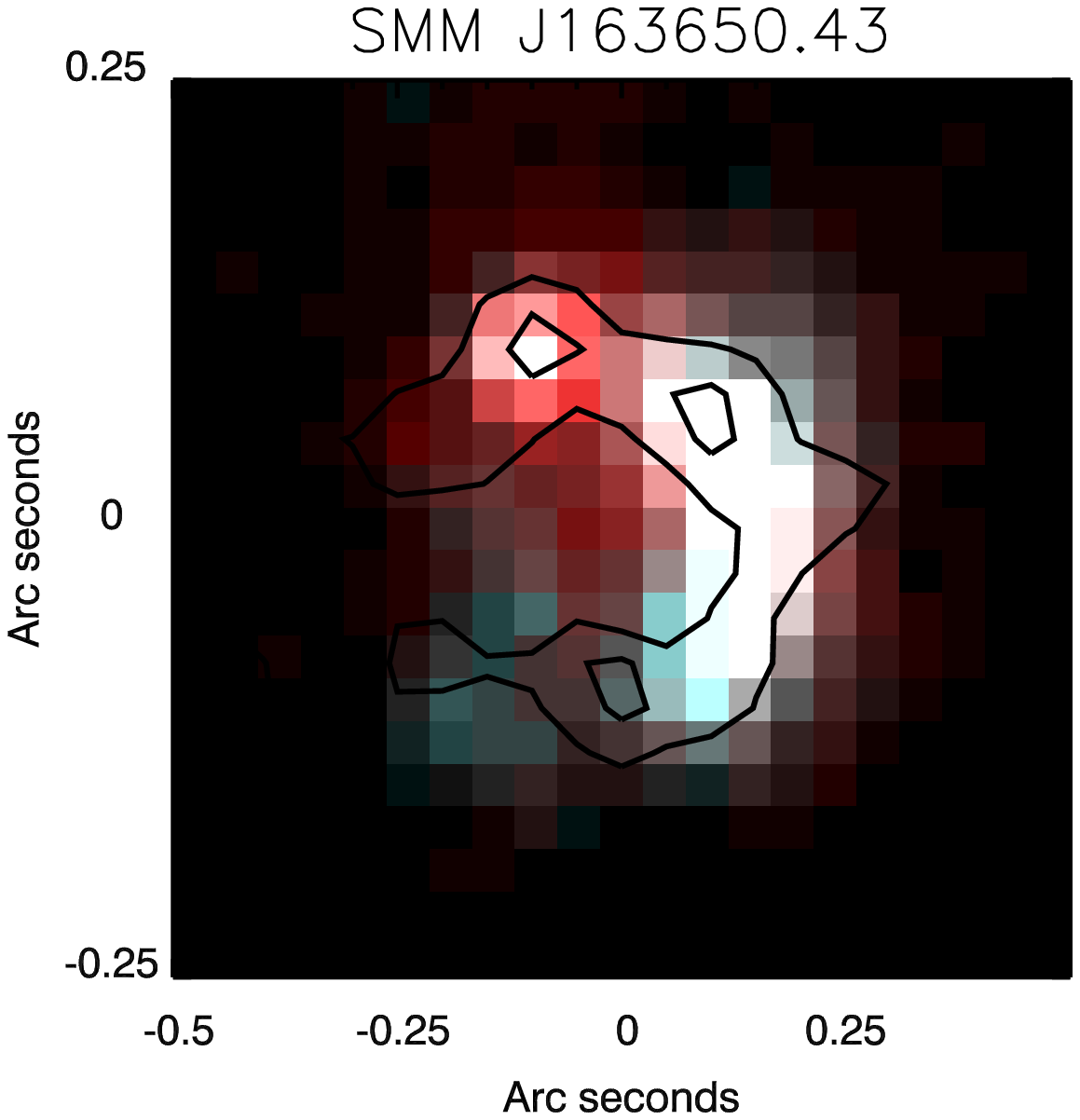}
\includegraphics[trim = 0mm -50mm 0mm 0mm, clip, scale=0.28, angle=0]{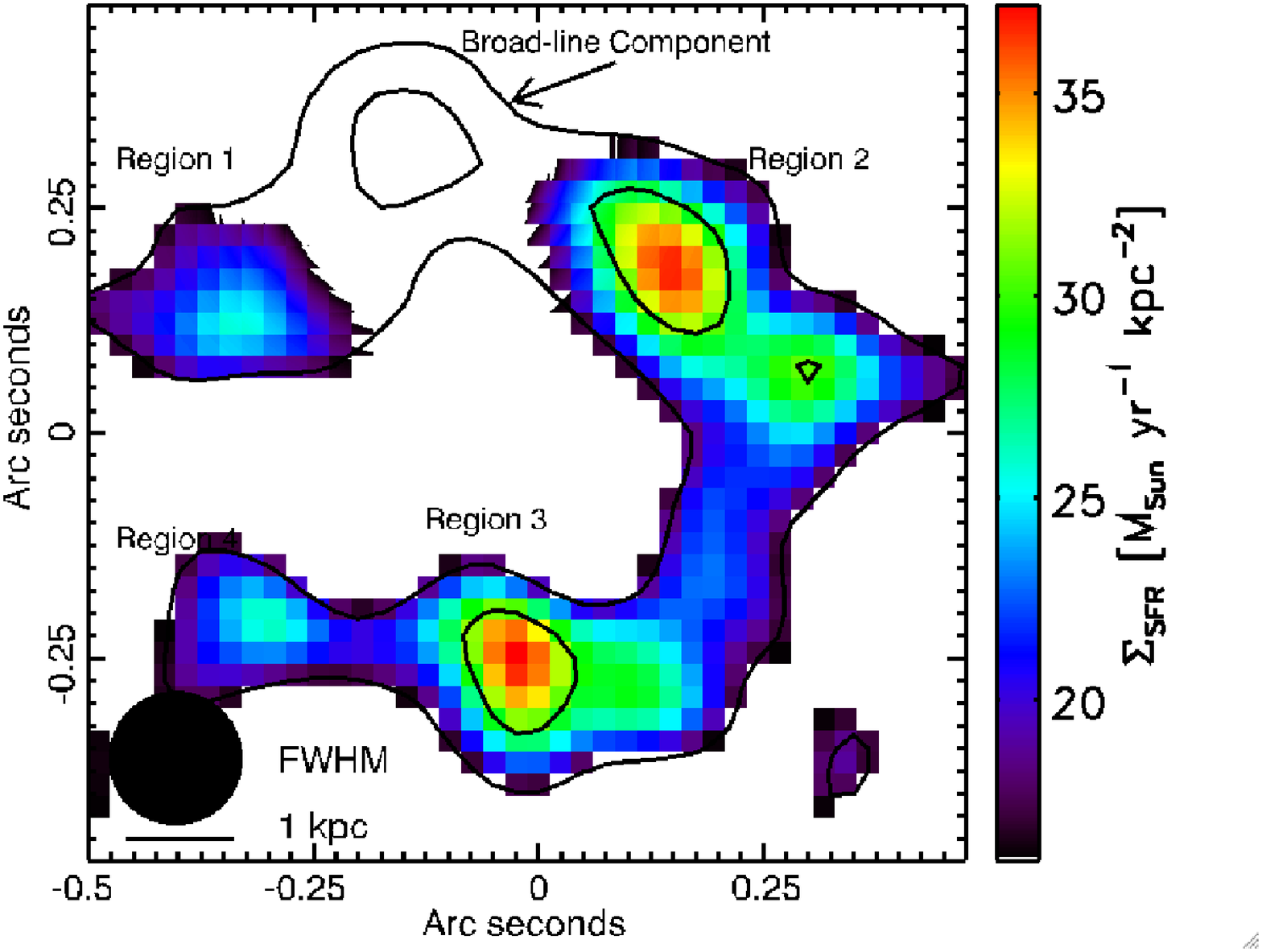}
\caption{Distribution of H$\alpha$ emission in SMM\,J163650.43, following the format of Fig.\,\ref{c3_15maps1}. The H$\alpha$ OSIRIS contours (inner/outer surface brightness contours corresponding to $\sim 20/ 8\sigma$), overlaid on a multicolor $HST$-ACS/NICMOS (B-/I-band) image, show H$\alpha$ line emission tracing the continuum-bright regions of the galaxy. We find broad H$\alpha$ emission coincident with the red knot in the HST image (see Fig.\,\ref{16t3maps2}), indicative of AGN activity. In addition to this, we distinguish H$\alpha$-bright clumps within the diffuse H$\alpha$ component. We divide this component into 4 distinct regions and study them individually.
\label{16t3maps1}}  
\end{figure*} 

%
%

\subsubsection{SMM\,J163650.43} \label{16t3results}

This galaxy was first detected by \citet{scott02} and \citet{ivison02}. Its broad-band near-IR emission has been studied in detail by \citet{smail03}. With a bolometric luminosity L$_{bol} = (3 \pm 2) \times 10^{13}$ L$_\odot$, this galaxy is unusually luminous even for an SMG (\citealt{chapman03}). Optical and near-IR spectroscopy revealed a redshift of $z \simeq 2.38$ \citep{chapman03}, unveiled narrow H$\alpha$ emission with an underlying broad H$\alpha$ component (FWHM$_{rest}~\simeq~310~\pm~50$ and $1750~\pm~240$~km~s$^{-1}$, respectively; \citealt{swinbank04}) and a high [OIII]/H$\beta$ line ratio characteristic of a Seyfert AGN (\citealt{smail03}). This galaxy has been resolved into different components, revealing a complex structure \citep{smail03, swinbank05, swinbank06}. In particular, based on recent seeing-limited Gemini-North NIFS observations, \citet{harrison12} report intricate extensions of narrow [OIII] emission ($100\lesssim$ FWHM $\lesssim500$ km s$^{-1}$) and two kinematically-distinct broad [OIII] emission lines (FWHM$\sim1200$ km s$^{-1}$). Furthermore, sub-arcsecond resolution ($\sim 0.25-0.5\arcsec$) CO observations of SMM\,J163650.43 undertaken by \citet{tacconi08} show a two-peaked CO profile emission restricted to an elliptical region with an intrinsic FWHM size of $0.8\pm0.2\arcsec\times0.4\pm0.3\arcsec$. 

\begin{figure*}
\centering
\includegraphics[trim = 0mm 0mm 0mm 0mm, clip,scale=0.6, angle=90]{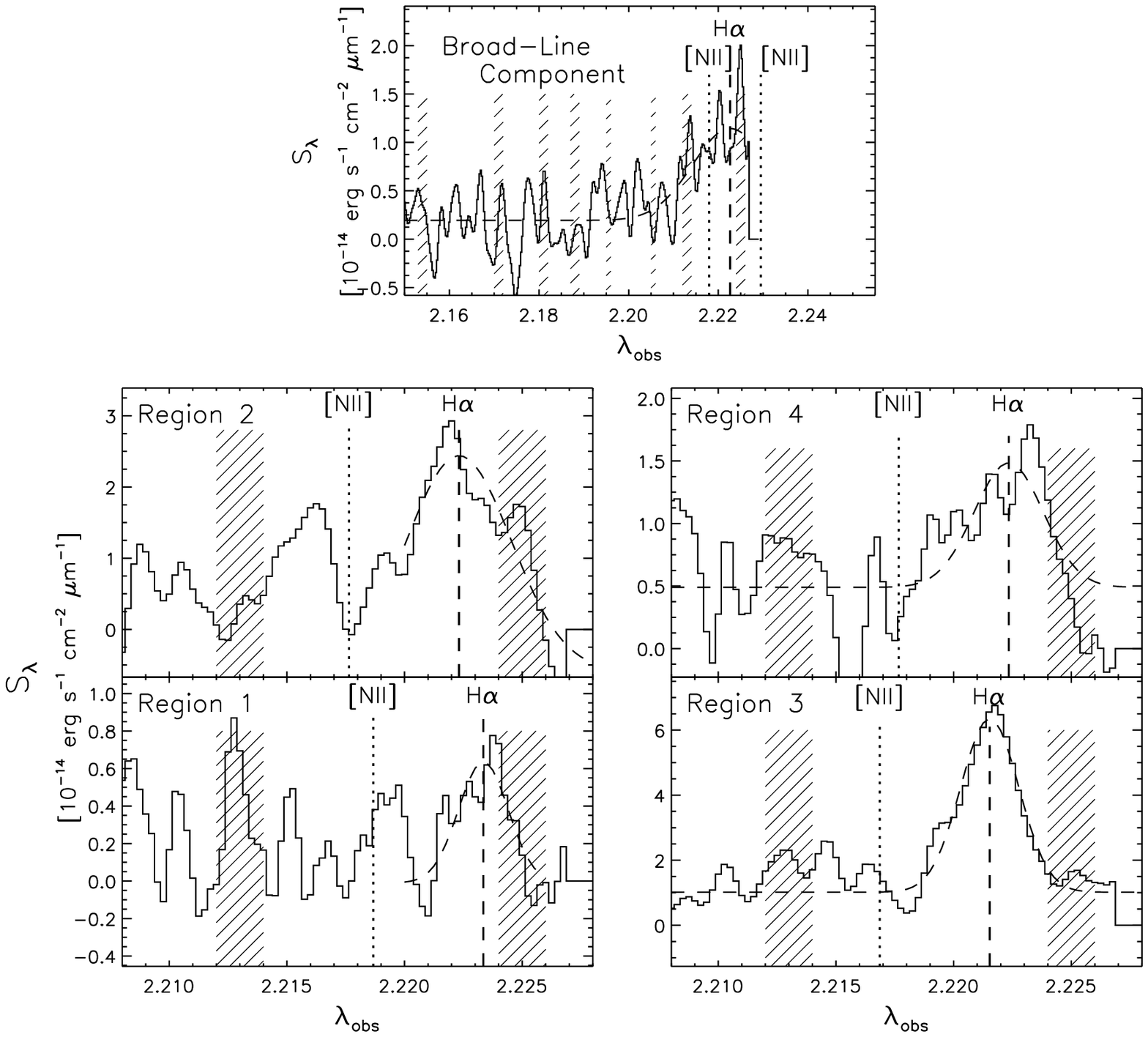}
\includegraphics[scale=0.26, angle=0]{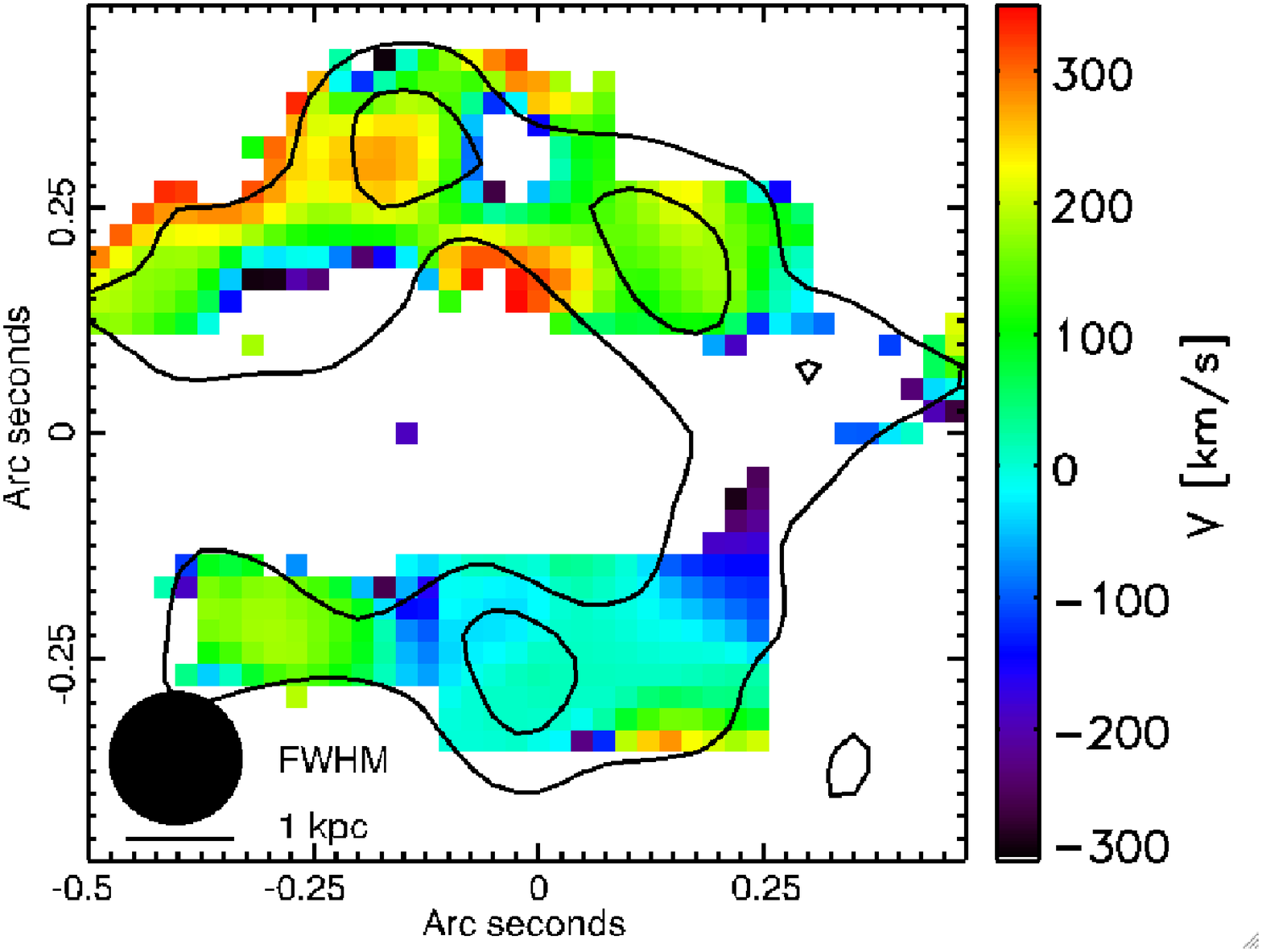}
\includegraphics[scale=0.26, angle=0]{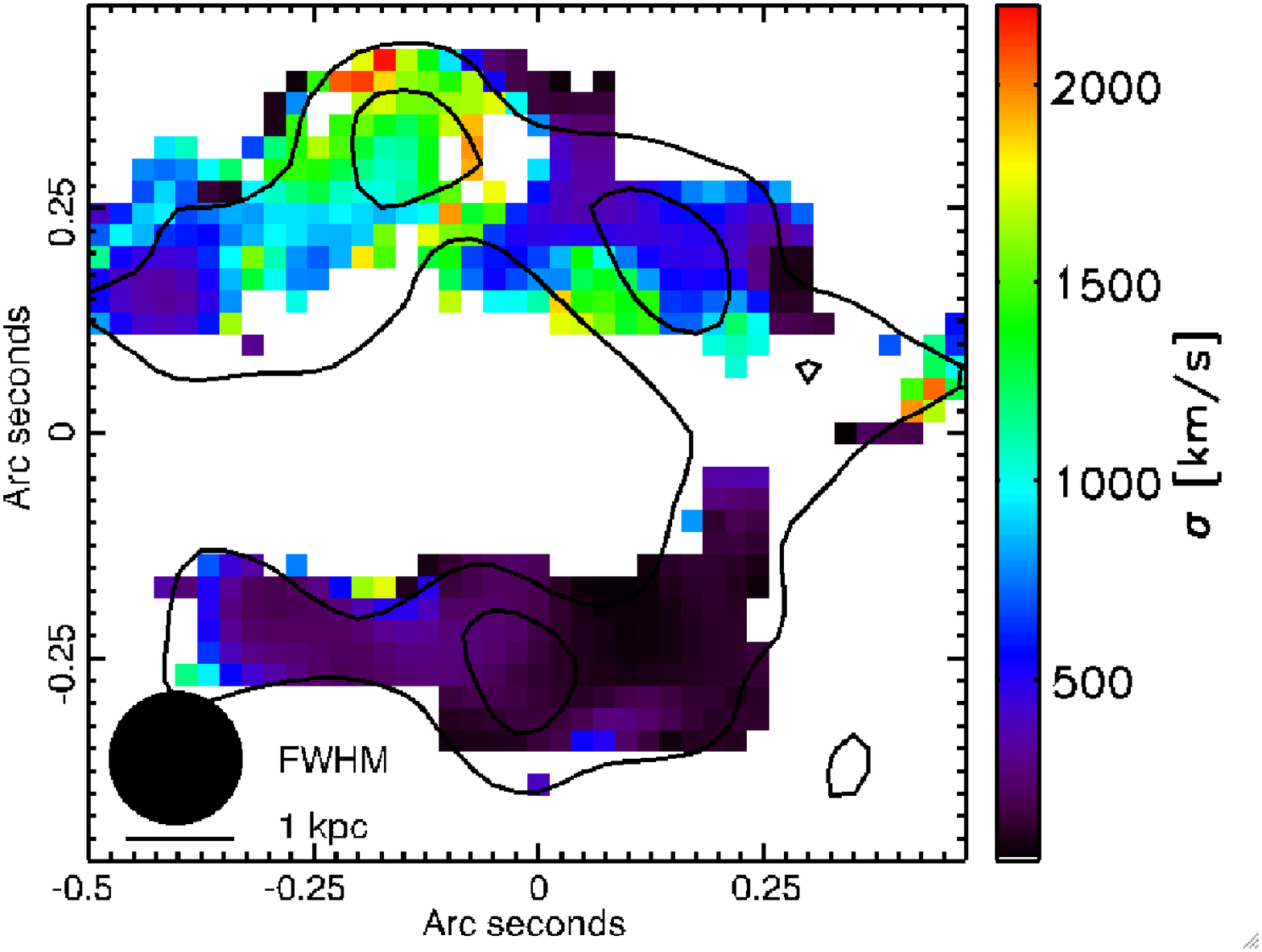}
\caption{H$\alpha$ kinematics in SMM\,J163650.43, following the format of Fig.\,\ref{c3_15maps2}. The 1D spectra suggest the presence of a broad H$\alpha$ line coincident with the region that appears as a red knot of continuum emission in Fig.\ref{16t3maps1}; this is indicative of AGN activity. 
\label{16t3maps2}}  
\end{figure*}

As shown in Fig.\,\ref{16t3maps1} we find H$\alpha$ emission in the shape of an arc, extending over $\sim5\times7$\,kpc$^2$ and tracing the continuum emission. The OSIRIS H$\alpha$ morphology agrees very well with the IFU results on this galaxy by \citet{swinbank05}, where the seeing-limited observations show 3 separate components corresponding to the northern, the western and the southern portions of the arc in Fig.\,\ref{16t3maps1}. Aided with AO, the finest OSIRIS lenslet-scale (0.05\arcsec) allows us to zoom further into the details of the inner regions of this galaxy: we distinguish 5 regions indicated in Fig.\,\ref{16t3maps1}. At a S/N$>10$, we distinguish broad H$\alpha$ emission associated with the north-eastern component coinciding with the red knot visible in the underlying HST image (see Figs.\,\ref{16t3maps1}-\ref{16t3maps2}). With a line-width of $\sigma\sim1000-1500$ km s$^{-1}$ (FWHM$\gtrsim3000$~km~s$^{-1}$; see Table\,\ref{ResultsTab}), we associate this emission to the broad-line region of an AGN. The two broad [OIII] components identified by \citet{harrison12} are spatially coincident with the broad H$\alpha$ region in our OSIRIS observations, but the velocity offsets ($\Delta v=\pm850$ km s$^{-1}$) associated to these components suggest that the [OIII] broad emission traces bi-polar outflows driven by the AGN.

We find narrower H$\alpha$ emission from Region 1 and from the western and southern regions of the arc (Regions 2, 3, 4), with $\sigma\lesssim300$ km s$^{-1}$ (FWHM$<700$ km s$^{-1}$; see Table\,\ref{ResultsTab}). Close to $\sim35$\% of the clump-integrated H$\alpha$ emission detected by OSIRIS lies within the extent of the broad-line region associated to the AGN (FWHM$\simeq0.2\arcsec \sim1.7$ kpc), while the remainder of the H$\alpha$ is spread to larger scales, with each of the 1-2 kpc-sized H$\alpha$ bright clumps accounting for $\sim3-30\%$. Based on the 1D spectra shown in Fig.\,\ref{16t3maps2}, we find velocity offsets corresponding to $\sim100-200$ km s$^{-1}$ between the different regions, in good agreement with previous findings by \citet{swinbank05}.

%
%

%
%
\begin{figure*}
\centering
\includegraphics[trim = 0mm 70mm 0mm 10mm, clip, scale=0.5, angle=90]{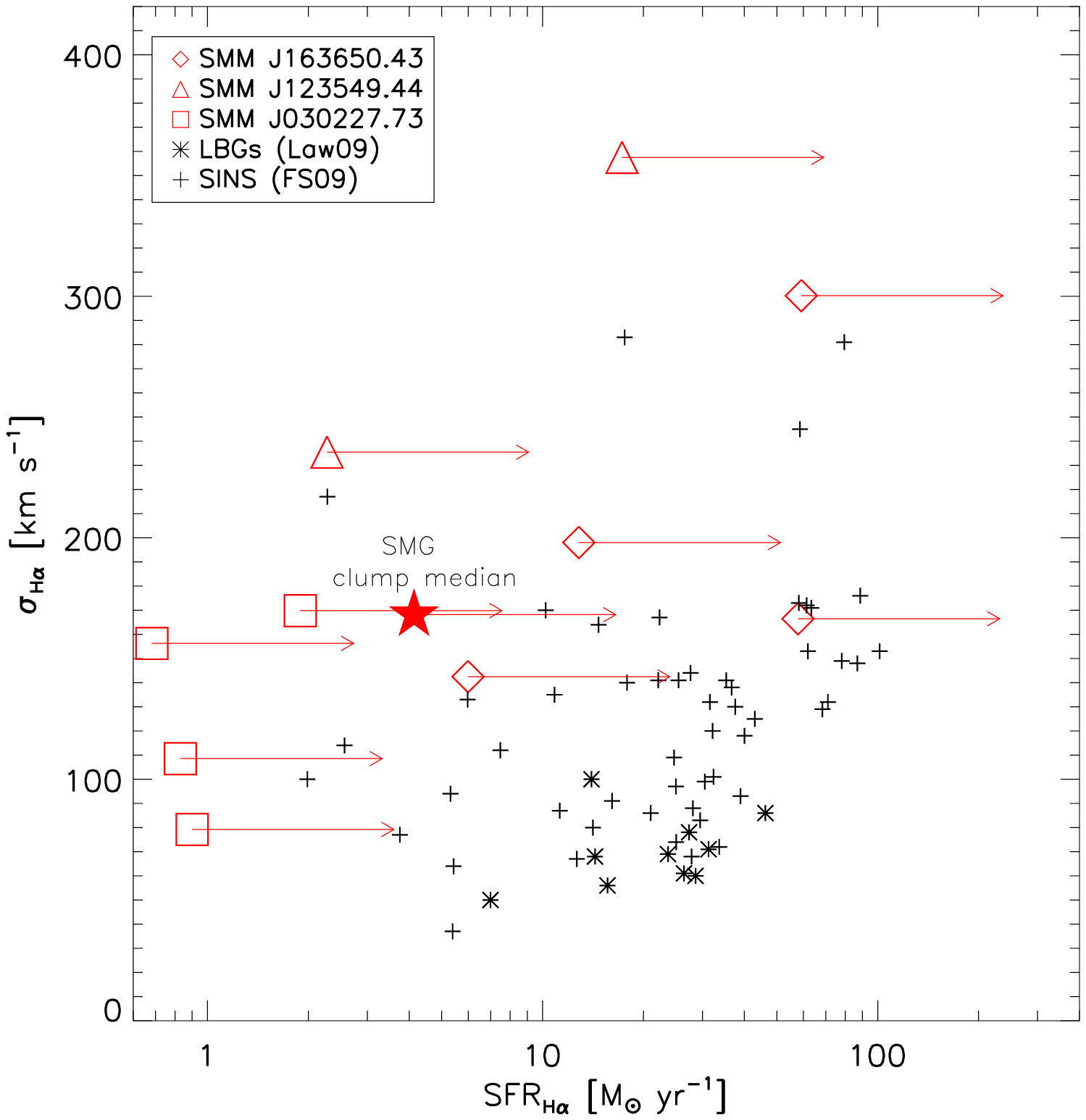}
\includegraphics[trim = 0mm 70mm 0mm 10mm, clip,scale=0.5, angle=90]{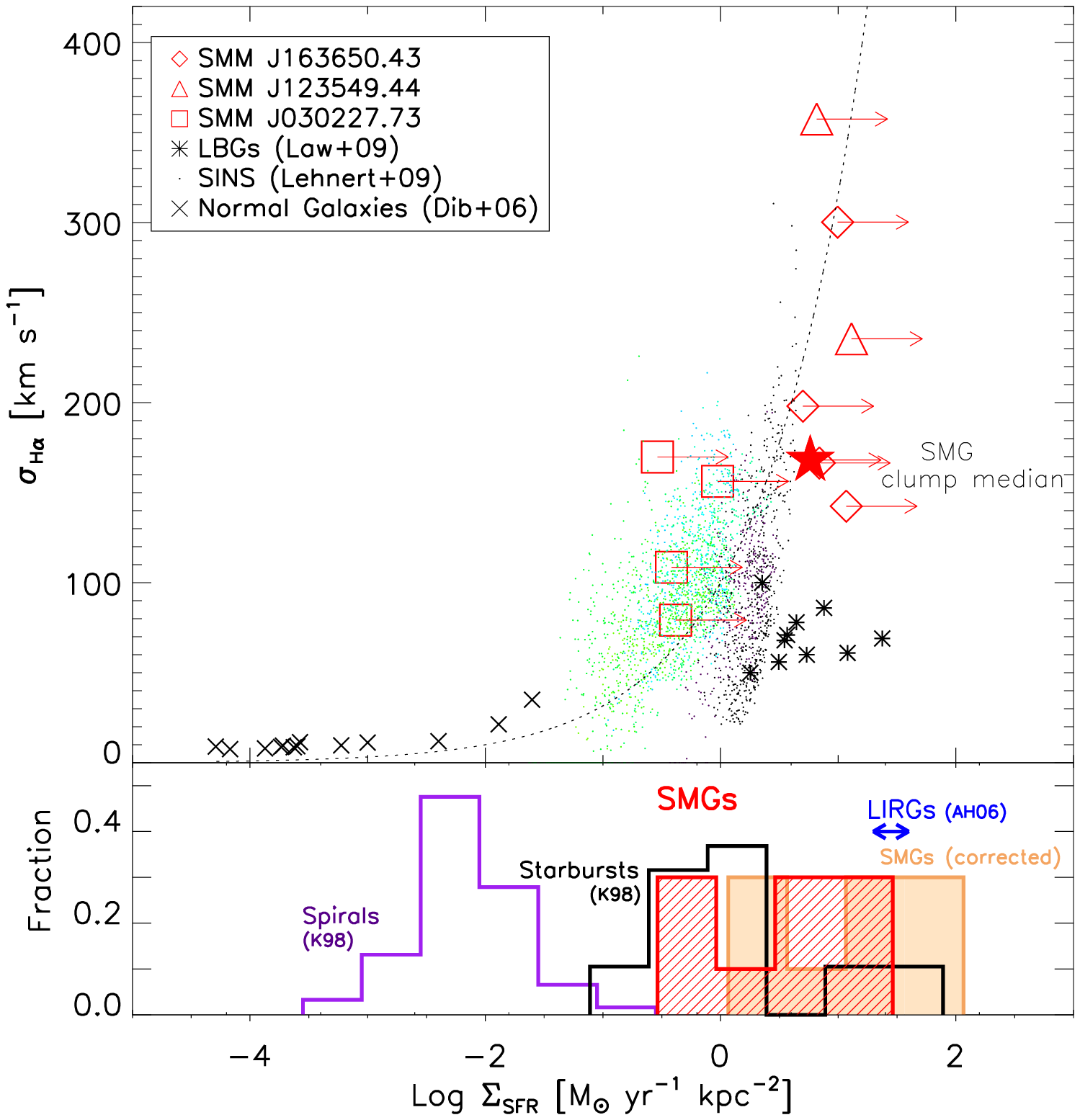}
\caption{SFR (left panel) and SFR surface density (right panel) as a function of velocity dispersions based on H$\alpha$ OSIRIS observations for individual SMG H$\alpha$-bright clumps (large symbols). The large filled star corresponds to the median value for the SMG clumps and the endpoints of the horizontal arrows indicate the result after correcting for extinction (see Section\,\ref{sfr} for details). For comparison we include values (not corrected for extinction) based on H$\alpha$ for local normal galaxies \citep{dib06} and at the high-redshift end, based on galaxy-integrated H$\alpha$ measurements for optically- and near-IR selected galaxies at $z\sim2-3$, including LBGs \citep{law09} and galaxies from the SINS sample (\citealt{forster-schreiber09}, FS09), and on spatially-resolved H$\alpha$ measurements on select SINS galaxies \citep{lehnert09}. The dashed curve on the right panel represents the functional form $\sigma \sim\sqrt (\epsilon \Sigma_{SFR})$ presented by \citet{lehnert09} to represent the coupling between the injection of mechanical energy by intense star-forming activity and the kinematics of the ionized gas, with $\epsilon=240$ representing 25\% coupling efficiency. To discuss SMG SFR intensities within a wider range of environments found in the low-redshift universe, we also include the distribution in SFR surface densities based on IR measurements for normal spirals and starbursts (\citealt{kennicutt98}, K08) and those for LIRGs based on Paschen-$\alpha$ measurements in comparably-extended star-forming regions ($\sim0.7-2$ kpc; \citealt{alonso-herrero06}, AH06). These figures show that SMGs clumps harbor very intense star-formation activity, occupying the high end of the $\Sigma_{\rm{SFR}}$ vs. $\sigma$ relation and sharing SFR intensities with that found in local extreme environments, such as nuclear starbursts and LIRGs.
\label{SFRD}}  
\end{figure*} 

%
%

\begin{deluxetable*}{rccc}
\centering
\tablewidth{0pt}
\tablecolumns{4}
\tabletypesize{\scriptsize}
\tablecaption{Summary of OSIRIS IFU Results} 
\tablehead{
 \colhead{}  &  \colhead{SMM J030227.73} &  \colhead{SMM J123549.44} &  \colhead{SMM J163650.43}   
\label{ResultsTab}}
\startdata
\tableline
S$_{H\alpha}$ [10$^{-17}$ erg s$^{-1}$ cm$^{-2}$]\,\tablenotemark{a} &   &  & \\
Region 1 &  0.9$\pm$0.4& 6.0$\pm$3.0 & 1.7$\pm$1.2 \\
Region 2 &  2.0$\pm$0.9 & 0.8$\pm$0.4 & 17.0$\pm$8.0 \\
Region 3 &  0.9$\pm$0.5 & $-$ & 16.5$\pm$3.0 \\
Region 4 &  0.7$\pm$0.6 & $-$ & 3.7$\pm$2.0 \\
Broad-line Region &  59.0$\pm$21.0 & 110.0$\pm$23.0 & 20.0$\pm$4.0 \\ \\
SFR$_{H\alpha}$ [M$_\odot$ yr$^{-1}$]\,\tablenotemark{b} &   &  & \\
Region 1 & 0.9 & 17.0 & 6.0 \\
Region 2 & 1.9 & 2.3 & 60.0 \\
Region 3 & 0.8 & $-$ & 58.0 \\
Region 4 & 0.7 & $-$ & 13.0 \\ \\
$\Sigma_{\rm{SFR}}$ [M$_\odot$ yr$^{-1}$ kpc$^{-2}$] &   &  & \\
Region 1 & 0.4 & 6.6 & 12.0 \\
Region 2 & 0.3 & 13.0 & 10.0 \\
Region 3 & 0.4 & $-$ & 7.0 \\
Region 4 & 0.9 & $-$ & 5.0 \\ \\
FWHM$_{\rm{rest}}$ [km s$^{-1}$] &  &  & \\
Region 1 & 190$\pm$60 & 840$\pm$340 & 330$\pm$160 \\ 
Region 2 & 400$\pm$130 & 550$\pm$190 & 700$\pm$220 \\ 
Region 3 & 260$\pm$100 &  $-$ & 390$\pm$50 \\ 
Region 4 & 370$\pm$230 &  $-$ & 470$\pm$170 \\ 
Broad-line Region & 3100$\pm$650 & 1260$\pm$210 &  2660$\pm$390 \\ 
\tableline
\enddata
\tablenotetext{a}{~Flux-calibrated H$\alpha$ flux (not corrected for extinction) from 1D spectra shown in Figs.\,\ref{c3_15maps2}, \ref{12t1maps2}, \ref{16t3maps2}.}
\tablenotetext{b}{~Based on \citet{kennicutt98}.}
\end{deluxetable*}
 
%
%

\subsection{Star formation rate surface density in SMGs}\label{sfr}

Based on \citet{kennicutt98} we derive a local SFR at each spatial pixel in our flux-calibrated H$\alpha$ images for each SMG target. Taking  into account the pixel scale of our observations (see Table\,\ref{Obstab}) we construct the SFR surface density ($\Sigma_{\rm{SFR}}$) maps shown in Figs.\,\ref{c3_15maps1}, \ref{12t1maps1}, \ref{16t3maps1}. The regions of broad H$\alpha$ emission, which we assume to be associated with AGN activity, have been subtracted following the approach described in Section\,\ref{maps} to cleanly eliminate the AGN contribution. The remaining narrow-line H$\alpha$ emission shown in these figures, likely associated to star formation, extends out to scales $>0.5-2\arcsec$ ($\sim4-16$ kpc). However, the bulk of this emission appears to be concentrated in multiple H$\alpha$ unresolved clumps, limited in size by the PSF FWHM$\sim0.10-0.25\,\arcsec$ ($1-2$ kpc at $z\sim2$). 

We note that although the near-IR provides us with a less-obscured insight to these SMG clumps than the optical, the presence of significant obscuration remains an important caveat \citep{takata06}. In the absence of spatially-resolved extinction maps, we make an estimated extinction correction based on the typical Balmer decrement found for SMGs: \citet{takata06} have shown that SMGs typically have observed Balmer decrements H$\alpha$/H$\beta~\sim~5-20$, corresponding to extinction levels in the range A$_v \sim 1-4$ with a median $<A_v> = 2.9 \pm 0.5$. Assuming a typical Balmer decrement $\sim 10$, we apply the corresponding reddening correction of $e^{\tau_{Balmer}} \sim 4$ to the measured H$\alpha$ luminosities. Note that this factor is significantly larger than the typical attenuation found for other high-redshift populations, such as the optically-selected LBGs, with attenuations of an average factor of 1.7 (\citealt{erb06}).  

Prior to extinction correction, SMG clumps have already large SFR surface densities with peak values $\Sigma_{\rm{SFR}} \sim 1- 50$ (see Figs.\,\ref{c3_15maps1}, \ref{12t1maps1}, \ref{16t3maps1}). After correcting for extinction in the manner we described, the peak SFR surface densities increase up to $\Sigma_{\rm{SFR}}\sim200\,$M$_\odot$ yr$^{-1} $kpc$^{-2}$  and the clump-wide SFRs reach values of up to 250 M$_\odot$ yr$^{-1}$. In Fig.\,\ref{SFRD} we show the total SFRs and SFR surface densities as a function of the line velocity dispersions based on the clump-wide spectral extractions shown in Figures \ref{c3_15maps2}, \ref{12t1maps2}, \ref{16t3maps2}, with the extinction-corrected values shown by the endpoints of horizontal arrows. For comparison, we include both kpc-scale and galaxy-wide measurements for optically- and near-IR selected galaxies at $z\sim2-3$ \citep{lehnert09, forster-schreiber09, law09} and for samples of low-redshift normal and starburst galaxies, compiled by \citet{dib06}. We also compare the distribution in SFR surface densities in our SMG regions with that in low-redshift samples of normal galaxies and circumnuclear starbursts from \citet{kennicutt98} and luminous infrared galaxies (LIRGs; L$_{\rm{8-1000\mu m}} \sim 10^{11}-10^{12}$L$_\odot$) from \citet{alonso-herrero06}.

The SMG clump-integrated SFRs extend over a large range of values (SFR$_{\rm{clump}}\sim1-60$M$_\odot$ yr$^{-1}$) and are similar to the galaxy-wide values that have been attributed to LBGs \citep{law09} and other high-redshift galaxies from the SINS sample (e.g., \citealt{forster-schreiber09}), particularly when extinction correction is taken into account (see left panel in Fig.\,\ref{SFRD}). It has been shown that a trend exists between $\Sigma_{\rm{SFR}}$ and H$\alpha$ velocity dispersion in these high redshift galaxy populations, where the most intense star-forming regions are also those with the highest velocity dispersions \citep{lehnert09, law09, forster-schreiber09}. On the right panel of Fig.\,\ref{SFRD} we show that the SMG clumps in our sample occupy the high end of this trend. Although the SMG clumps in this sample present a wide range of extinction-corrected SFR and $\Sigma_{\rm{SFR}}$ values (SFR$\sim3-250$ M$_\odot$ yr$^{-1}$, $\Sigma_{\rm{SFR}}\sim1-50\,$M$_\odot$yr$^{-1}$ kpc$^{-2}$), the average values that characterize the star formation activity in these ($<$SFR$>\sim60$ M$_\odot$ yr$^{-1}$, $<\Sigma_{\rm{SFR}}>\sim10\,$M$_\odot$yr$^{-1}$ kpc$^{-2}$) is similar in intensity -- though higher by a factor of 2--5 in some cases -- to that found in other high-redshift populations (e.g., $<$SFR$>_{\rm{LBGs}} \sim 25$M$_\odot$ yr$^{-1}$, $<\Sigma_{\rm{SFR}}>_{\rm{LBGs}}  \sim 4$M$_\odot$ yr$^{-1}$ kpc$^{-2}$, \citealt{law09}). However, the H$\alpha$ velocity dispersions are significantly different, likely associated to the different mass ranges that these different galaxy populations correspond to. Assuming that $\sigma_{\rm{H\alpha}}$ traces gas dynamics within virialized clouds and considering a typical clump size of 1 kpc, we determine clump dynamical masses in the range of M$_{\rm{clumps}}=\sigma^2R/G \sim 1-30 \times 10^{9}$ M$_\odot$. These clump masses are $\sim1-2$ orders of magnitude larger than the kpc-scaled stellar clumps uncovered by \citet{forster-schreiber11} in the optically-selected  sample of $z\sim2$ SINS star-forming galaxies (M$_{ \rm{clumps}}\sim 0.1-8 \times 10^{9}$ M$_\odot$). 

\citet{lehnert09} argue that the relationship between SFR surface density and the ionized gas velocity dispersion is a consequence of star-formation self-regulation, where the mechanical energy of the starburst itself controls the local dynamics of the ionized gas. Following \citet{lehnert09}, this relationship can be represented by a functional form involving the intensity of star formation with the velocities of the ionized gas to represent the energy injected into the ISM surrounding the star-forming regions (see curve in Fig.\,\ref{SFRD}). Within this context, the high velocity dispersions observed in our SMG clumps are likely not the sole result of larger dynamical masses. They are likely a combination of differences in dynamical mass ranges and a reflection of the high pressures sustained by high local surface brightnesses, which in turn are likely due to the higher gas densities in SMGs (e.g., \citealt{harris10}).

Compared to galaxies at low redshift, the right panel of Fig.\,\ref{SFRD} shows that SMG clumps have $\Sigma_{\rm{SFR}}$ values $>3$ orders of magnitude higher than those found in normal spirals (\citealt{kennicutt98, dib06}), but which fall well within the range of starburst galaxies \citep{kennicutt98}. If we correct for extinction the  $\Sigma_{\rm{SFR}}$ in SMG clumps are shifted at the most within a factor of a few from the range occupied by the typical local circumnuclear starbursts, similar to the range found in low-redshift LIRGs \citep{alonso-herrero06}. 

%
%

\section{Discussion}\label{discuss}

Near-IR integral field spectroscopic observations have shown that H$\alpha$ emission in SMGs often hint to the presence of multiple spatially-distinct galactic-scale subcomponents within the central $\sim1\arcsec$ \citep{swinbank06, nesvadba07}. With LGS-AO OSIRIS we have been able to peer into these central regions, allowing us to zoom in further than these previous seeing-limited integral field observations. 

%
%

\subsection{Internal Dynamics: SMGs are Kinematically-Disturbed Systems}\label{dyn}

In all three SMGs of our sample we identify regions with distinct properties: regions with narrow H$\alpha$ emission where star formation is likely taking place; and regions with broad H$\alpha$ emission, likely dominated by AGN activity (keeping in mind that outflows may also play a role: e.g., \citealt{shapiro08, shapiro09, newman12}). We find that the narrow-H$\alpha$ emission is concentrated in multiple clumps unresolved on $\sim1-2$\,kpc scales, asymetrically distributed around the AGN. These clumps suggest either one of the following scenarios: (1) these H$\alpha$-bright regions correspond to distinct components in a disturbed system, as would be found in a merging system; or, (2) we are picking out the high surface-brightness peaks of H$\alpha$ emission in an underlying lower surface-brightness plane of emission -- perhaps even a disk, as has been found in a sub-sample of massive optically-selected star-forming galaxies at $z\sim2-3$ (\citealt{forster-schreiber06, law07b, law09, forster-schreiber11}).

We find velocity offsets of $\sim$ {\it few} $\times100$ km s$^{-1}$ between distinct galactic-scale regions (see Figs.\,\ref{c3_15maps2}, \ref{12t1maps2}, \ref{16t3maps2}).  These velocity offsets could be explained by invoking a merger scenario (see also \citealt{engel10}), thus strengthening the conclusion derived from deep rest-frame optical HST-imaging that SMGs are disturbed systems, likely corresponding to mergers (\citealt{smail98, smail04, ivison10, swinbank10, aguirre12}). Within this context the distinct components revealed by OSIRIS could potentially be associated with the remnants of the pre-merger galaxies, that have not fully coalesced and still retain significant intrinsic velocities with respect to each other. This is the kinematic structure expected for the intermediate stages of a merging system (e.g., \citealt{cox06, hayward11}). We note that this is in agreement with \citet{greve05}, where they discuss the likelihood that the double-peaked CO profiles observed in a sample of 18 SMGs correspond to either a disk or a merger. Taking into account estimates of gas masses and source sizes, they conclude that such high surface mass density would imply dynamical instabilities too powerful for a disk to survive. They thus conclude that the double-peaked nature of the CO lines traces distinct gas-rich components undergoing a merger or arise from a disk collapsing under gravitational instability. 

On the other hand, the H$\alpha$-bright clumps of emission that OSIRIS picks out may correspond to distinct clumps within an underlying clumpy disk structure. This would be similar to the clumpy disks identified in face-on galaxies at somewhat lower redshifts ($z\sim 0.5-2.0$: \citealt{elmegreen04a, elmegreen04b}) and the more recently identified stellar clumps in optically- and near-IR selected star-forming galaxies at $z\sim2$ by \citet{shapiro09}, \citet{forster-schreiber11}, \citet{genzel11} and \citet{newman12}. Recent studies have indeed suggested the presence of disk structures in a large fraction of SMGs based on deep broad-band imaging K-band data (e.g., \citealt{targett11}). Although these studies are unable to directly trace the kinematic structure of these objects, they may point to the possibility that SMGs may correspond to a diverse population presenting a varied range of kinematic structures.

Considering that the broad-H$\alpha$ emission in our SMGs identifies a super-massive black hole (SMBH) likely at the dynamical center of system, and the large masses and extreme star-forming properties of the SMG clumps that are in turn asymmetrically distributed around the AGN, it is unlikely that these all reside in a regular potential well structure. The merger interpretation is likely the most accurate scenario for the SMGs in our sample, although the final test of whether an underlying disk structure is present will come from studies of the cold gas at the high spatial resolutions with ALMA.

%
%

\subsection{Extended starbursts at high redshift}\label{gas}

We find compact unresolved regions (FWHM$\sim1-3$ kpc) of broad-line H$\alpha$ emission that contribute a significant fraction ($\sim30-90\%$) to the galaxy-wide H$\alpha$ emission. However, fainter narrow-line H$\alpha$ emission extends over large spatial scales of $4-16$\,kpc (see Figs.\,\ref{c3_15maps1}, \ref{12t1maps1}, \ref{16t3maps1}), with 1-2 kpc-sized H$\alpha$-bright clumps indicating regions of particularly intense star-formation (see linewidth maps in Figs.\,\ref{c3_15maps2}, \ref{12t1maps2}, \ref{16t3maps2}). Spatially-extended sizes for the diffuse H$\alpha$ emission have also been identified in the seeing-limited IFU observations presented by \citet{swinbank06} and \citet{alaghband-zadeh12}, who report resolved nebular emission on scales from $4-11$\,kpc for a combined sample of 12 SMGs. 

Evidence of spatially-extended emission in SMGs has also been found at longer wavelengths. In high-resolution observations of a range of CO transitions \citep{engel10, ivison11} and radio continuum \citep{biggs08}, linear sizes in the range of $\sim1-16$\,kpc have been identified based on median angular-averaged FWHM sizes, with sample median sizes of $\sim5-7$\,kpc. High-resolution radio continuum observations by \citet{chapman04} also unveil spatial extensions with a linear diameter out to $\gtrsim 10$\,kpc, while a high-resolution far-IR study by \citet{younger10} reveal far-IR emission extending out to spatial scales in the range of $\sim5-8$ kpc. 

The similarity in IR luminosities between local ultra-luminous infrared galaxies (ULIRGs; L$_{\rm{8-1000\mu m}} > 10^{12}$L$_\odot$) and high-redshift SMGs, as well as the presence of AGN signatures in both populations have naturally motivated direct comparisons and discussions of their possible correspondence within the formation and evolution scenario of today's most massive spheroids. To a somewhat still controversial degree, the sizes of SMGs also play a role in this debate. High-resolution studies of the CO and nebular emission in SMGs have concluded that SMGs have {\it compact} sizes \citep{tacconi06, tacconi08}, although often with companions with an average separation of $\sim8\pm2$ \citep{alaghband-zadeh12}, while other studies have demonstrated the presence of more {\it extended} spatial extensions \citep{chapman04, tecza04, nesvadba07, biggs08, hailey-dunsheath10}. The former argues for a similarity with local ULIRGs, which are found to be compact with sizes $\sim 1$\,kpc (in the mid-IR, \citealt{charmandaris02, diaz-santos10b}; and in CO, \citealt{downes98, bryant99}); the latter argues for a sharp difference with local ULIRGs. However, an important role in this controversy has been played by the tenuous definition of what is {\it compact} with respect to what is {\it extended}; particularly in the case of a clumpy system, as we find in SMGs. 

In the light of our OSIRIS observations, together with these previous findings, it is clear that the SMG population displays a range in spatial extensions that are often significantly larger than those of local ULIRGs. On the other hand, SMGs share similar star formation intensities with LIRGs (see Fig.\,\ref{SFRD}), which are still within the range of extremely active environments, yet within a somewhat lower IR luminosity bin than ULIRGs.  LIRGs have also been found to display rest-frame optical line emission over large spatial scales ($\sim3-7$ kpc; \citealt{alonso-herrero06}) as well as mid-IR extra-nuclear emission \citep{diaz-santos10a, diaz-santos10b}, both indicative of extended star-formation. We emphasize that evidence for extended star formation at high redshifts is not limited to SMGs. Some of the brightest and most massive LBGs do appear to have comparable H$\alpha$ spatial extensions ($R_{H\alpha} \sim 2-7$\,kpc; \citealt{forster-schreiber06}).

Our results indicate that the SMG clumps in our sample have high surface densities of star formation activity, close to those found in local extreme environments, such as in circumnuclear starbursts and luminous infrared galaxies. However, considering the much greater spatial extents found for these SMGs ($\sim8-16$\,kpc) in comparison to the 1-kpc sized nuclear starbursts \citep{kennicutt98}, SMGs appear to be undergoing this intense activity on much larger spatial scales. 

%
%

\section{CONCLUSIONS}\label{conclusions}

The advent of integral field units (IFUs) in large-aperture optical telescopes has pushed detailed galaxy kinematic studies out to the high-redshift realm. We present the first integral-field spectroscopic Laser Guide Star Adaptive Optics (LGS-AO) observations of submillimeter galaxies (SMGs). The OSIRIS instrument on Keck with LGS-AO allow us to separate spatial and spectral information at sufficiently high spatial resolutions to determine SFRs from H$\alpha$-bright regions uncontaminated by the broad-line emission associated to AGN activity and to explore the internal dynamics of these complex systems. Our main results are the following:
\newline
\newline
\noindent $\bullet$ We spatially distinguish between the compact broad-H$\alpha$ emission (FWHM$\sim0.2-0.4\arcsec$, corresponding to $\sim2-3$ kpc) associated with an AGN and the multiple 1-2 kpc-sized narrow-H$\alpha$ clumps of emission associated with star formation, asymetrically distributed around the AGN. This had remained unachievable in prior long-slit spectroscopic studies and seeing-limited IFU observations of SMGs. 
\newline
\newline
\noindent $\bullet$ We find that the H$\alpha$ emission arising from the broad-line component may sometimes contribute up to $\sim$90\% to the total H$\alpha$ emission enclosed in the bright SMG clumps, while the contribution from the individual stellar clumps varies from $1-30\%$, with a median value of $\sim3\%$. These contributions do not translate directly to an AGN and star-formation contribution to the total luminous output, since inhomogeneous dust extinction remains unconstrained at this stage.
\newline
\newline
\noindent $\bullet$  We do not find any indication of ordered global motion within our targets, as would be found in rotationally-supported disks.  We find relative velocities of a {\it few} $\times$ 100 km s$^{-1}$ between the stellar clumps and the AGN in our systems, suggesting that these SMGs do not represent regular potential well structures, but are more likely in an intermediate merging phase. However, the final test of whether an underlying disk structure is present will come from detailed studies of the cold molecular gas at the high spatial resolutions possible with ALMA.
\newline
\newline
\noindent $\bullet$ SMGs seem to display high SFR surface densities ($\Sigma_{\rm{SFR}}$) similar to those found in the most extreme local environments, such as circumnuclear starbursts and IR-luminous galaxies. However, because the narrow-line H$\alpha$ emission spreads over large spatial extensions $\sim 4-16$\,kpc, this sets them in sharp contrast to local ULIRGs. All of these results taken together suggest that the submillimeter phase denotes a short-lived flaring-up of large spatial extension across these systems that rapidly depletes the available gas through intense star formation. 

\acknowledgements We thank the referee for her/his useful comments and suggestions. We also thank David R. Law and Shelley Wright for helpful and insightful discussions on the treatment and analysis of OSIRIS observations. We are also grateful to the Keck support team for the fantastic on-site help in obtaining these observations, in particular to Al Conrad, Randy Campbell, Hien Tran, David LeMignant, Jim Lyke and Christine Melcher. The data presented herein were obtained at the W.M. Keck Observatory, which is operated as a scientific partnership among the California Institute of Technology, the University of California and the National Aeronautics and Space Administration. The Observatory was made possible by the generous financial support of the W.M. Keck Foundation. The authors wish to recognize and acknowledge the very significant cultural role and reverence that the summit of Mauna Kea has always had within the indigenous Hawaiian community.  We are most fortunate to have the opportunity to conduct observations from this mountain. KMD was supported by an NSF Astronomy and Astrophysics Postdoctoral Fellowship under award AST-0802399. AWB was supported by the NSF under award AST-0909159. IRS acknowledges support from the Royal Society. This research has made use of the NASA/ IPAC Infrared Science Archive, which is operated by the Jet Propulsion Laboratory, California Institute of Technology, under contract with the National Aeronautics and Space Administration.


\begin{thebibliography}

\bibitem[Aguirre et al.(2012)]{aguirre12} Aguirre, P. et al. 2012, \apj, submitted

\bibitem[Alaghband-Zadeh et al.(2012)]{alaghband-zadeh12} Alaghband-Zadeh, S., et al. 2012, \mnras, 424, 2232

\bibitem[Alexander et al.(2005)]{alexander05}Alexander, D. M., Bauer, F. E., Chapman, S., Smail, I., Blain, A.,  Brandt, W. N., Ivison, R. 2005, \apj, 632, 736 (A05)

\bibitem[Alexander et al.(2008)]{alexander08} Alexander, D.~M., et al.\ 2008, AJ, 135, 1968 

\bibitem[Alonso-Herrero et al.(2006)]{alonso-herrero06} Alonso-Herrero, A., Rieke, G.~H., Rieke, M.~J., Colina, L., P{\'e}rez-Gonz{\'a}lez, P.~G., \& Ryder, S.~D.\ 2006, \apj, 650, 835 

\bibitem[Barger et al.(1998)]{barger98} Barger, A.~J., Cowie, L.~L., Sanders, D.~B., Fulton, E., Taniguchi, Y., Sato, Y., Kawara, K., \& Okuda, H.\ 1998, \nat, 394, 248 

\bibitem[Barger, Cowie \& Sanders(1999)]{barger99}Barger, A. J., Cowie, L. L., \& Sanders, D. B. 1999, \apj, 518, L5

\bibitem[Basu-Zych et al.(2009)]{basu-zych09} Basu-Zych, A.~R., et al.\ 2009, \apjl, 699, L118 

\bibitem[Bertoldi et al.(2000)]{bertoldi00} Bertoldi, F., et al.\ 2000, A\&A, 360, 92 

\bibitem[Biggs \& Ivison(2008)]{biggs08} Biggs, A.~D., \& Ivison, R.~J.\ 2008, \mnras, 385, 893 

\bibitem[Biggs et al.(2010)]{biggs10} Biggs, A.~D., Younger, J.~D., \& Ivison, R.~J.\ 2010, \mnras, 408, 342 

\bibitem[Blain et al.(2002)]{blain02}Blain, A., Smail, I., Ivison, R., Kneib, J.-P., Frayer, D. T.   2002, Phys. Rep., 369, 111B

\bibitem[Borys et al.(2003)]{borys03}Borys, C., Chapman, S., Halpern, M., \& Scott, D. 2003, \mnras, 344, 385

\bibitem[Bournaud et al.(2009)]{bournaud09} Bournaud, F., Elmegreen, B.~G., \& Martig, M.\ 2009, \apjl, 707, L1 

\bibitem[Bournaud et al.(2011)]{bournaud11} Bournaud, F., et al.\ 2011, \apj, 730, 4 

\bibitem[Bryant \& Scoville(1999)]{bryant99} Bryant, P.~M., \& Scoville, N.~Z.\ 1999, \aj, 117, 2632 

\bibitem[Chapman et al.(2003)]{chapman03}Chapman, S., Blain, A.,  Ivison, R., Smail, I. 2003, Nature, 422, 695

\bibitem[Chapman et al.(2004)]{chapman04} Chapman, S.~C., Smail, I., Windhorst, R., Muxlow, T., \& Ivison, R.~J.\ 2004, \apj, 611, 732 

\bibitem[Chapman et al.(2005)]{chapman05}Chapman, S., Blain, A., Smail, I., Ivison, R. 2005, \apj, 622, 772 (C05)

\bibitem[Charmandaris et al.(2002)]{charmandaris02} Charmandaris, V., et al.\ 2002, \aap, 391, 429 

\bibitem[Conselice et al.(2004)]{conselice04} Conselice, C.~J., et al.\ 2004, \apjl, 600, L139 

\bibitem[Coppin et al.(2005)]{coppin05} Coppin, K., Halpern, M., Scott, D., Borys, C., \& Chapman, S.\ 2005, MNRAS, 357, 1022 

\bibitem[Cowie et al.(2002)]{cowie02}Cowie, L. L., Barger, A. J., \& Kneib, J.-P. 2002, \aj, 123, 2197

\bibitem[Cox et al.(2006)]{cox06} Cox, T.~J., Dutta, S.~N., Di Matteo, T., Hernquist, L., Hopkins, P.~F., Robertson, B., \& Springel, V.\ 2006, \apj, 650, 791 

\bibitem[Davies(2007)]{davies07} Davies, R.~I.\ 2007, \mnras, 375, 1099 

\bibitem[Dekel et al.(2009)]{dekel09} Dekel, A., Sari, R., \& Ceverino, D.\ 2009, \apj, 703, 785 

\bibitem[D{\'{\i}}az-Santos et al.(2010a)]{diaz-santos10a} D{\'{\i}}az-Santos, T., Alonso-Herrero, A., Colina, L., Packham, C., Levenson, N.~A., Pereira-Santaella, M., Roche, P.~F., \& Telesco, C.~M.\ 2010, \apj, 711, 328 

\bibitem[D{\'{\i}}az-Santos et al.(2010b)]{diaz-santos10b} D{\'{\i}}az-Santos, T., et al.\ 2010, \apj, 723, 993 

\bibitem[Dib et al.(2006)]{dib06} Dib, S., Bell, E., \& Burkert, A.\ 2006, \apj, 638, 797 

\bibitem[Dickinson et al.(2003)]{dickinson03} Dickinson, M., Papovich, C., Ferguson, H.~C., \& Budav{\'a}ri, T.\ 2003, \apj, 587, 25 

\bibitem[Downes \& Solomon(1998)]{downes98} Downes, D., \& Solomon, P.~M.\ 1998, \apj, 507, 615 

\bibitem[Eales et al.(1999)]{eales99}Eales, S., Lilly, S., Gear, W., Dunne, L., Bond, J.R., Hammer, F., Le F\`{e}vre, O., \&  Crampton, D. 1999, \apj, 515, 518

\bibitem[Elmegreen et al.(2004a)]{elmegreen04a} Elmegreen, D.~M., Elmegreen, B.~G., \& Sheets, C.~M.\ 2004, \apj, 603, 74 

\bibitem[Elmegreen et al.(2004b)]{elmegreen04b} Elmegreen, D.~M., Elmegreen, B.~G., \& Hirst, A.~C.\ 2004, \apjl, 604, L21 

\bibitem[Elmegreen \& Elmegreen(2005)]{elmegreen05} Elmegreen, B.~G., \& Elmegreen, D.~M.\ 2005, \apj, 627, 632 

\bibitem[Engel et al.(2010)]{engel10} Engel, H., et al.\ 2010, \apj, 724, 233 

\bibitem[Erb et al.(2003)]{erb03} Erb, D.~K., Shapley, A.~E., Steidel, C.~C., Pettini, M., Adelberger, K.~L., Hunt, M.~P., Moorwood, A.~F.~M., \& Cuby, J.-G.\ 2003, \apj, 591, 101 

\bibitem[Erb et al.(2006)]{erb06} Erb, D.~K., Steidel, C.~C., Shapley, A.~E., Pettini, M., Reddy, N.~A., \& Adelberger, K.~L.\ 2006, \apj, 647, 128 

\bibitem[F{\"o}rster Schreiber et al.(2006)]{forster-schreiber06} F{\"o}rster Schreiber, N.~M., et al.\ 2006, \apj, 645, 1062 

\bibitem[F{\"o}rster Schreiber et al.(2009)]{forster-schreiber09} F{\"o}rster Schreiber, N.~M., et al.\ 2009, \apj, 706, 1364 

\bibitem[F{\"o}rster Schreiber et al.(2011)]{forster-schreiber11} F{\"o}rster Schreiber, N.~M., Shapley, A.~E., Genzel, R., et al.\ 2011, 
\apj, 739, 45 

\bibitem[Genzel et al.(2003)]{genzel03} Genzel, R., Baker, A.~J., Tacconi, L.~J., Lutz, D., Cox, P., Guilloteau, S., \& Omont, A.\ 2003, \apj, 584, 633 

\bibitem[Genzel et al.(2011)]{genzel11} Genzel, R., et al.\ 2011, \apj, 733, 101 

\bibitem[Greve et al.(2005)]{greve05} Greve, T.~R., et al.\ 2005, \mnras, 359, 1165 

\bibitem[Gon{\c c}alves et al.(2010)]{goncalves10} Gon{\c c}alves, T.~S., et al.\ 2010, \apj, 724, 1373 

\bibitem[Harris et al.(2010)]{harris10} Harris, A.~I., Baker, A.~J., Zonak, S.~G., et al.\ 2010, \apj, 723, 1139 

\bibitem[Hailey-Dunsheath et al.(2010)]{hailey-dunsheath10} Hailey-Dunsheath, S., Nikola, T., Stacey, G.~J., Oberst, T.~E., Parshley, S.~C., Benford, D.~J., Staguhn, J.~G., \& Tucker, C.~E.\ 2010, \apjl, 714, L162 

\bibitem[Hainline et al.(2009)]{hainline09} Hainline, L.~J., Blain, A.~W., Smail, I., Frayer, D.~T., Chapman, S.~C., Ivison, R.~J., \& Alexander, D.~M.\ 2009, \apj, 699, 1610 

\bibitem[Hainline et al.(2011)]{hainline11} Hainline, L.~J., Blain, A.~W., Smail, I., Alexander, D. M., Armus, L., Chapman, S.~C., \& Ivison, R.~J.\ 2011, \apj, 740, 96 

\bibitem[Harrison et al.(2012)]{harrison12} Harrison, C.~M., Alexander, D.~M., Swinbank, A.~M., et al.\ 2012, \mnras, 426, 1073 

\bibitem[Hayward et al.(2011)]{hayward11} Hayward, C.~C., Kere?, D.,  Jonsson, P., Narayanan, D., Cox, T.~J., Hernquist, L. \ 2011, \apj, 743, 159

\bibitem[Hughes et al.(1998)]{hughes98} Hughes, D.~H., et al.\ 1998, \nat, 394, 241 

\bibitem[Immeli et al.(2004)]{immeli04} Immeli, A., Samland, M., Gerhard, O., \& Westera, P.\ 2004, \aap, 413, 547 

\bibitem[Ivison et al.(2002)]{ivison02} Ivison, R.~J., et al.\ 2002, \mnras, 337, 1 

\bibitem[Ivison et al.(2010)]{ivison10} Ivison, R.~J., Smail, I., Papadopoulos, P.~P., Wold, I., Richard, J., Swinbank, A.~M., Kneib, J.-P., \& Owen, F.~N.\ 2010, \mnras, 404, 198 

\bibitem[Ivison et al.(2011)]{ivison11} Ivison, R.~J., Papadopoulos, P. P., Smail, I., Greve, T. R., Thomson, A. P., Xilouris, E. M.,  Chapman, S. C.\ 2011, \mnras, 412, 1913 

\bibitem[Jones et al.(2010)]{jones10} Jones, T.~A., Swinbank, A.~M., Ellis, R.~S., Richard, J., \& Stark, D.~P.\ 2010, \mnras, 404, 1247 

\bibitem[Kennicutt(1998)]{kennicutt98} Kennicutt, R.~C., Jr.\ 1998, \apj, 498, 541 

\bibitem[Laird et al.(2010)]{laird10} Laird et al. 2010, \mnras, 401, 2763

\bibitem[Larkin et al.(2006)]{larkin06} Larkin, J., et al.\ 2006, Proceedings of the SPIE, 6269, 62691A

\bibitem[Law et al.(2007b)]{law07b} Law, D.~R., Steidel, C.~C., Erb, D.~K., Larkin, J.~E., Pettini, M., Shapley, A.~E., \& Wright, S.~A.\ 2007, \apj, 669, 929 

\bibitem[Law et al.(2007a)]{law07a} Law, D.~R., Steidel, C.~C., Erb, D.~K., Pettini, M., Reddy, N.~A., Shapley, A.~E., Adelberger, K.~L., \& Simenc, D.~J.\ 2007, \apj, 656, 1 


\bibitem[Law et al.(2009)]{law09} Law, D.~R., Steidel, C.~C., Erb, D.~K., Larkin, J.~E., Pettini, M., Shapley, A.~E., \& Wright, S.~A.\ 2009, \apj, 697, 2057 

\bibitem[Lehnert et al.(2009)]{lehnert09} Lehnert, M.~D., Nesvadba, N.~P.~H., Tiran, L.~L., Matteo, P.~D., van Driel, W., Douglas, L.~S., Chemin, L., \& Bournaud, F.\ 2009, \apj, 699, 1660 

\bibitem[Lilly et al.(1999)]{lilly99} Lilly, S.~J., Eales, S.~A., Gear, W.~K.~P., Hammer, F., Le F{\`e}vre, O., Crampton, D., Bond, J.~R., \& Dunne, L.\ 1999, ApJ, 518, 641 

\bibitem[Melbourne et al.(2009)]{melbourne09} Melbourne, J., et al.\ 2009, \aj, 137, 4854 

\bibitem[Men{\'e}ndez-Delmestre et al.(2009)]{md09} Men{\'e}ndez-Delmestre, K., et al.\ 2009, \apj, 699, 677

\bibitem[Nesvadba et al.(2007)]{nesvadba07} Nesvadba, N.~P.~H., et al.\ 2007, \apj, 657, 725 

\bibitem[Newman et al.(2012)]{newman12} Newman, S., et al.\ 2012, \apj, 752, 111 

\bibitem[Overzier et al.(2010)]{overzier10} Overzier, R.~A., Heckman, T.~M., Schiminovich, D., Basu-Zych, A., Gon{\c c}alves, T., Martin, D.~C., \& Rich, R.~M.\ 2010, \apj, 710, 979 

\bibitem[P{\'e}rez-Gonz{\'a}lez et al.(2008)]{perez-gonzalez08} P{\'e}rez-Gonz{\'a}lez, P.~G., et al.\ 2008, \apj, 675, 234 

\bibitem[Pope et al.(2008)]{pope08} Pope, A., et al.\ 2008, \apj, 675, 1171 

\bibitem[Scott et al.(2002)]{scott02}Scott, S. et al. 2002,  \mnras, 331, 817

\bibitem[Shapiro et al.(2008)]{shapiro08}Shapiro, K. et al. 2008,  \apj, 682, 231

\bibitem[Shapiro et al.(2009)]{shapiro09}Shapiro, K. et al. 2009,  \apj, 701, 955

\bibitem[Smail et al.(1997)]{smail97}Smail, I., Ivison, R., Blain, A.1997, \apj, 490L, 5S

\bibitem[Smail et al.(1998)]{smail98} Smail, I., Ivison, R.~J., Blain, A.~W., \& Kneib, J.-P.\ 1998, \apj, 507, L21 

\bibitem[Smail et al.(2003)]{smail03} Smail, I., Chapman, S.~C., Ivison, R.~J., Blain, A.~W., Takata, T., Heckman, T.~M., Dunlop, J.~S., \& Sekiguchi, K.\ 2003, \mnras, 342, 1185 

\bibitem[Smail et al.(2004)]{smail04} Smail, I., Chapman, S.~C., Blain, A.~W., \& Ivison, R.~J.\ 2004, \apj, 616, 71 

\bibitem[Stark et al.(2008)]{stark08} Stark, D.~P., Swinbank, A.~M., Ellis, R.~S., Dye, S., Smail, I.~R., \& Richard, J.\ 2008, \nat, 455, 775 

\bibitem[Swinbank et al.(2004)]{swinbank04}Swinbank, A. M., Smail, I., Chapman, S., Blain, A.,     Ivison, R., Keel, W. C. 2004, \apj, 617, 64

\bibitem[Swinbank et al.(2005)]{swinbank05} Swinbank, A.~M., et al.\ 2005, \mnras, 359, 401 

\bibitem[Swinbank et al.(2006)]{swinbank06} Swinbank, A.~M., Chapman, S.~C., Smail, I., Lindner, C., Borys, C., Blain, A.~W., Ivison, R.~J., \& Lewis, G.~F.\ 2006, \mnras, 371, 465 

\bibitem[Swinbank et al.(2010)]{swinbank10} Swinbank, A.~M., et al.\ 2010, \mnras, 405, 234 

\bibitem[Tacconi et al.(2006)]{tacconi06} Tacconi, L.~J., et al.\ 2006, \apj, 640, 228 

\bibitem[Tacconi et al.(2008)]{tacconi08} Tacconi, L.~J., et al.\ 2008, \apj, 680, 246 

\bibitem[Takata et al.(2006)]{takata06} Takata, T., Sekiguchi, K., Smail, I., et al.\ 2006, \apj, 651, 713 

\bibitem[Targett et al.(2011)]{targett11} Targett, T.~A., Dunlop, J.~S., McLure, R.~J., et al.\ 2011, \mnras, 412, 295 

\bibitem[Tecza et al.(2004)]{tecza04} Tecza, M., et al.\ 2004, \apjl, 605, L109 

\bibitem[van Dam et al.(2006)]{vandam06} van Dam, M.~A., et al.\ 2006, PASP, 118, 310 

\bibitem[Wardlow et al.(2011)]{wardlow11} Wardlow, A., et al.\ 2011, \mnras, 415, 1479

\bibitem[Wei{\ss} et al.(2009)]{weiss09} Wei{\ss}, A., et al.\ 2009, \apj, 707, 1201 

\bibitem[Webb et al.(2003a)]{webb03a} Webb, T.~M., et al.\ 2003a, \apj, 587, 41 

\bibitem[Webb et al.(2003b)]{webb03b}Webb, T. M. A., Lilly, S., Clements, D. L., Eales, S., Yun, M., Brodwin, M., Dunne, L., \& Gear, W. 2003b, \apj, 597, 680

\bibitem[Wizinowich et al.(2006)]{wizinowich06} Wizinowich, P.~L., et al.\ 2006, PASP, 118, 297 

\bibitem[Wisnioski et al.(2012)]{wisnioski12} Wisnioski, E. et al.\ 2012, \mnras, 422, 3339

\bibitem[Wright et al.(2009)]{wright09} Wright, S.~A., Larkin, J.~E., Law, D.~R., Steidel, C.~C., Shapley, A.~E., \& Erb, D.~K.\ 2009, \apj, 699, 421 

\bibitem[Younger et al.(2007)]{younger07} Younger, J. D., et al.\  2007,  \apj, 671, 1531 

\bibitem[Younger et al.(2010)]{younger10} Younger, J.~D., et al.\  2010, \mnras, 407, 1268 


\end{thebibliography}
\end{document}